\documentclass[final,12pt]{article}
\usepackage[utf8]{inputenc}
\usepackage{anysize}
\marginsize{2cm}{2cm}{2cm}{2cm} 

\usepackage{fancyhdr} 
\usepackage{lipsum} 
\usepackage{color} 
\usepackage{bm} 
\usepackage{amsfonts,amsthm,amsmath,amssymb} 
\usepackage{latexsym,graphicx,epstopdf} 
\usepackage{etoolbox} 
\usepackage{hyperref}

\renewcommand{\a}{{\bf a}}

\newcommand{\n}{{\bf n}}

\renewcommand{\t}{{\bf t}}
\renewcommand{\u}{{\bf u}}

\newcommand{\x}{{\bf x}}

\newcommand{\zero}{{\bf 0}}

\newcommand{\bb}{{\bf B}}

\newcommand{\ff}{{\bf F}}

\newcommand{\ii}{{\bf I}}

\newcommand{\btau}{{\bm\tau}}

\newcommand{\bsigma}{{\bm\sigma}}

\newcommand{\bdiv}{{\bf div}}

\newcommand{\transpose}{\texttt{t}}

\title{The energy of muscle contraction. I. Tissue force and deformation during isometric contractions\thanks{This work was partially supported by the Natural Sciences and Engineering Research Council of Canada and the Comisi\'on Nacional de Investigatici\'on Cient\'ifica y Tecnol\'ogica of Chile.}}

\author{James M. Wakeling\thanks{Department of Biomedical Physiology and Kinesiology, Simon Fraser University, Burnaby, BC, Canada}{\,\,\,\thanks{Department of Mathematics, Simon Fraser University, Burnaby, BC, Canada}}\quad Stephanie A. Ross\footnotemark[2]\quad David S. Ryan\footnotemark[2]\\ Bart Bolsterlee\thanks{Neuroscience Research Australia, Randwick, New South Wales, Australia}\quad 
Ryan Konno\thanks{Department of Physics and Astronomy, University of British Columbia, Canada}\\ Sebasti\'an Dom\'inguez\footnotemark[3]{\,\,\,\thanks{Corresponding author: 
\href{mailto:domingue@sfu.ca}{domingue@sfu.ca}}}\quad Nilima Nigam\footnotemark[3]}

\begin{document}

\maketitle

\begin{abstract}
During contraction the energy of muscle tissue increases due to energy from the hydrolysis of ATP.
This energy is distributed across the tissue as strain-energy potentials in the contractile elements,
strain-energy potential from the 3D deformation of the base-material tissue (containing cellular and
ECM effects), energy related to changes in the muscle’s nearly incompressible volume and external
work done at the muscle surface. Thus, energy is redistributed through the muscle’s tissue as it
contracts, with only a component of this energy being used to do mechanical work and develop
forces in the muscle’s line-of-action. Understanding how the strain-energy potentials are redistributed
through the muscle tissue will help enlighten why the mechanical performance of whole muscle in its
line-of-action does not match the performance that would be expected from the contractile elements
alone. Here we demonstrate these physical effects using a 3D muscle model based on the finite
element method. The tissue deformations within contracting muscle are large, and so the mechanics
of contraction were explained using the principles of continuum mechanics for large deformations.
We present simulations of a contracting medial gastrocnemius muscle, showing tissue deformations
that mirror observations from MRI-based images. This paper tracks the redistribution of strain-
energy potentials through the muscle tissue during isometric contractions, and shows how fibre
shortening, pennation angle, transverse bulging and anisotropy in the stress and strain of the muscle
tissue are all related to the interaction between the material properties of the muscle and the action of
the contractile elements.

\end{abstract}


{\bf Keywords}: muscle, energy, finite element model, MRI, contraction, tissue, deformation, 3D
\vspace{.25cm}


\section{Introduction}\label{section:intro}
Most of our understanding of muscle function and performance comes from measurements at
small scales such as sarcomeres, single fibres and small muscles. Additionally, muscle contraction
data have typically been determined when muscle is fully active, changes length at constant velocity,
and considers forces and length changes in only the longitudinal direction. By comparison, we know
much less about how whole, large muscles contract, particularly when they are not fully active and
contract with varying velocities. Yet, these are exactly the conditions that we may want to understand
in order to understand healthy muscle function, and the impairments that arise from injury, disuse
and disease. Understanding how the contractile elements interact with the tissue properties of the
whole muscle, how deformations may arise in all three dimensions during contraction, and how the
dynamics of muscle size influences whole muscle performance may result in muscle behaviours that
are not intuitive from the understanding of single fibre function alone. The purpose of this series of
papers is to consider contractile mechanisms that are relevant at the whole muscle level, and how
these influence the design and performance of skeletal muscle.

Muscles change shape and develop forces when they contract. These effects are typically
assumed to occur along the line-of-action of the muscle, however, shape changes and forces can
occur in all three dimensions. For example, as a muscle shortens then it must increase in girth, or
cross-sectional area, in order to maintain its volume (Zuurbier and Huijing, 1993; Böl et al., 2013;
Randhawa and Wakeling, 2015). Additionally, as a muscle expands in cross-section it will tend to
push outwards as transverse forces develop. Indeed, transverse expansions have been reported from
early studies, where contracting muscle bulged to fill glass tubes (Swammerdam, 1758, source:
Cobb, 2002), to more recent studies where muscle bulging has been reported in both animal
(Brainerd and Azizi, 2005; Azizi et al., 2008) and human studies (Randhawa et al., 2013; Dick and
Wakeling, 2017). Transverse forces and deformations have also been recorded by muscles lifting
weights when they bulge (Siebert et al., 2012; Ryan et al., 2019), which is akin to lifting your body
by tensing your glutes whilst you are seated.

The 3D shape changes of a muscle are important to its function (Azizi et al., 2008). The
forces that muscle fibres actively develop decrease the faster they shorten (Hill, 1938), and thus
processes that affect fibre shortening velocity will also affect their force. As the fibres shorten then
they must expand in girth to maintain their volume, making the fibres press on each other in a
transverse direction. In pennate muscle this transverse expansion is accommodated by the fibres
rotating to greater pennation angles (Alexander, 1983; Maganaris et al., 1998), or expanding in either
of the two transverse directions (Wakeling and Randhawa, 2014; Randhawa and Wakeling, 2018).
The increases in pennation angle result in lower fibre shortening velocity allowing the fibres to
develop greater forces, in a process known as muscle belly gearing (Wakeling et al., 2011). The
forces developed by whole muscle affect how it changes shape and can cause gearing to vary (Dick
and Wakeling, 2017), with this variable gearing favouring velocity output at low loads and force
output against high loads (Azizi et al., 2008). Changes to the 3D shape of muscle therefore influence
the deformations and speeds at which the fibres shorten, and consequently affect whole muscle
forces.

Transverse forces acting at the surface of muscle are also important to muscle function. When
groups of muscles within anatomical compartments contract together, their transverse bulging causes
the muscles to press on each other, and this results in lower forces being generated by the collective
group of muscles than is possible by the sum of the muscle forces if they are isolated (Fontana et al.,
2018). In a similar manner, when compressive forces are applied to the transverse surfaces of
contracting muscle, the muscle forces generated along their line-of-action decreases (Siebert et al.,
2012, 2016, 2018; Ryan et al., 2019), and the deformations of the fibres, changes in pennation angle
and belly gearing are also affected (Wakeling et al., 2013; Ryan et al., 2019). Additionally, the
tendency for a muscle to bulge can cause the muscle to do external work in a transverse direction:
generating a transverse force that that can lift a weight (Siebert et al., 2012). Contracting muscle thus
develops and reacts to transverse forces acting on its surface, and when the surface deforms in the
transverse directions, this will additionally result in transverse work being done.

This paper will consider the effect of work and energy on muscle contractions. Mechanical
work is the amount of energy transferred by a force. For clarity in this paper, the term work will be
used to describe mechanical work at the surface of the muscle, whereas the term energy will be used
to describe the internal energy within the muscle. This internal energy is a strain-energy, which is
energy stored by a system undergoing deformation. At the whole muscle level, any process that
redistributes energy into a transverse direction will detract from the energy that can be used to
generate mechanical work in the longitudinal line-of-action of the muscle. Currently, we know
relatively little about how energy redistributes within the whole muscle structure, and how this
redistribution of energy relates the transverse to the longitudinal work done by the contracting
muscle. However, energy redistribution within muscle may have important implications to both the
mechanical and metabolic function of a muscle (Williams et al., 2012; Roberts et al., 2019).
Surprisingly, these energetic considerations have barely been incorporated into our current
understanding of whole muscle function.

This is the first of a series of papers in which we explore how the redistribution of energy
within muscle affects its mechanical and metabolic function during contraction, and we use these
energetic mechanisms to demonstrate how whole muscle function is not simply due to the behaviour
of individual contractile elements, but rather emerges from the mechanics of the whole 3D muscle
structure. In this current paper we present a mechanistic framework for quantifying the energy
redistribution and describe how strain-energy is related to the stretch and shortening of the muscle
fibres, to the 3D shape and deformations of the whole muscle, and to the forces developed during
muscle contraction. We extend this analysis in two companion papers to identify how transverse
forces and compression affect the longitudinal force of contracting muscle (Ryan et al., 2020) in its
line-of-action, and how the mechanical and metabolic cost of accelerating the inertial mass of the
muscle tissue lessens the mechanical performance and efficiency during dynamic contractions (Ross
et al., 2020).


\section{Approaches}\label{section:approaches}
Experimental and modelling studies have reported local variations in tissue deformations
within muscle (Pappas et al., 2002; Higham et al., 2007; Hodson-Tole et al., 2016), and these
variations can be explained by the internal mechanics of the muscle fibres and surrounding tissue
(Blemker et al., 2005; Rahemi et al., 2014). Thus, how a muscle’s tissue deforms during contraction
depends on the general structural and material properties of the muscle, rather than on the particular
features of the muscle’s surface geometry. However, a wide range of muscle sizes, shapes and
architectures exist (Wickiewicz et al., 1983; Lieber and Fridén, 2000), so while the same physical
principles govern the internal mechanics and behaviour of muscle, the actual tissue deformations and
stresses that develop during contraction also depend on muscle shape and architecture (Gans and
Bock, 1965; Lieber and Fridén, 2000). Hence, in this study we compare deformations of muscle tissue for geometries that span a range of pennation angles and cross-sectional areas to distinguish
general principles that do not rely on specific features of muscle shape.

The premise in this study is that strain-energy redistributes through the muscle tissue resulting
in changes to the force and external work of whole muscle. The maximum work from an active
sarcomere is given by the area under its force-length curve when it shortens very slowly along its
entire range of motion so that its contractile forces are close to their maximum isometric value at
each instant (Weis-Fogh and Alexander, 1977), giving a maximum strain energy-density of
approximately $1.5\cdot 10^5$ Jm$^{-3}$ , where the strain-energy density is the strain-energy for a given volume
of muscle tissue. The maximum muscle work possible would be approximately equal to the product
of the work from each sarcomere and the number of sarcomeres in the muscle, or equivalently the
product of the strain energy-density from the sarcomere and the volume of the muscle tissue. Here we
additionally compare blocks of muscle with different shapes and architectures, but the same initial
volumes so that we can evaluate how strain-energy is redistributed within them independently from
the effect of muscle size, or effectively the number of sarcomeres.

The mechanics of whole muscle contraction depend on many factors such as the geometry of
the muscle and properties of the tissue, and different models have been evaluated to explain how
individual factors influence contractile performance. However, here we present a general modelling
approach, using the principle of minimum total energy (Liu \& Quek, 2014), to explain many of these
different effects in one framework. Previously, muscle shape changes have been related to belly
gearing and shortening velocities using both 2D and 3D geometrical models (eg. Maganaris et al.,
1998; Azizi et al., 2008; Randhawa and Wakeling, 2015); transmission of forces and deformations
between transverse and longitudinal directions have been investigated with studies that used fluid and
hydrostatic models and experiments (Sleboda and Roberts, 2017; Azizi et al., 2017); and lumped
parameter models have accounted for tissue mass, accelerations and the mechanical cost of inactive
tissue (Günther et al., 2012; Ross et al., 2016, 2018b). However, the physical principles that relate
muscle shape and force should emerge from the complex interactions between the contractile
elements, the material properties of the tissue and the 3D structure of the muscle and not rely on
specific explanations for distinct examples. Modelling muscle as a fibre-reinforced composite
biomaterial allows the principles of continuum mechanics and the finite element method (FEM) to be
applied to this problem (Meier and Blickhan, 2000; Johansson et al., 2000; Yucesoy et al., 2002;
Oomens et al., 2003; Blemker et al., 2005; Röhrle and Pullman, 2007; Böl and Reese, 2008; Rahemi
et al., 2014): in this approach tissue deformations are associated with an energy function, usually
called the strain-energy function. The strain-energy function describes all the active, passive and
incompressibility behaviours of the muscle tissue, allowing these models to track the redistribution of
strain-energy potentials within the tissue. Thus, such FEM models are ideal for evaluating the
redistribution of energy within a contracting muscle.

In this paper we use a FEM model of muscle that we previously developed (Rahemi et al.,
2014, 2015; Ross et al., 2018b), but with a number of numerical and computational refinements. Both
muscle and aponeurosis tissue are modelled as fibre-reinforced composite biomaterials using the
principles of continuum mechanics. For the muscle tissue, the fibres in the model represent the
myofilaments that develop both active and passive forces (from the actomyosin cross-bridges and
titin molecules, respectively). These model fibres are non-linear actuators, with their forces being
calculated using a Hill-type modelling approach (e.g. Zajac, 1989). The fibres develop active forces
that increase with the activation level and their orientations are specified at each point, allowing the
pennation angle to be calculated. The material properties of the muscle are modelled as base material
(combining both intracellular and extracellular effects), and the whole muscle tissue is considered as nearly-incompressible. The aponeurosis tissue is also fibre-reinforced, but here the model fibres
represent collagen fibres within the aponeurosis that are aligned in the longitudinal direction and are
tangential to the aponeurosis sheets. Similar to the muscle, the aponeurosis has its own base material
properties and nearly incompressible constraints. Both the muscle and aponeurosis tissues are thus
transversely isotropic. In this paper we quantify the energy state of the different elements within the
model (the contractile strain-energy potential from the fibres, the base material strain-energy
potential, and the volumetric strain-energy potential that penalizes volume changes at each element),
and track the redistribution of energy between these elements as the isometric contractions progress.
Here we evaluate the deformations of the medial gastrocnemius between our modelled and MRI
results, and we quantify the redistribution of energy that occurred within a block of muscle in the
medial gastrocnemius, and across a series of additional blocks with varying geometry and
architecture.

\section{Methods}\label{section:methods}
In this paper we present a parallel modelling and experimental study to evaluate the changes in
internal energy during isometric muscle contraction. We model blocks of muscle with different sizes,
shapes and pennation angles to determine how these features affect the strain-energy, deformations
and forces of the muscles. To assess how valid these modelled effects are to whole muscle
contractions we compare the model outputs to the outputs of a block of muscle within the
experimentally-measured geometry of the medial gastrocnemius muscle. Additionally, we validate
the deformations of the medial gastrocnemius that are predicted by the model with experimentally
measured deformations of the muscle surface geometry and the internal fibre pennation angle.

\subsection{Finite element model}\label{section:fem}
\subsubsection{Formulation}\label{section:formulation}
We modelled the muscle tissue as a three-dimensional and nearly incompressible fibre-reinforced
composite material. While the model is transversely isotropic, the presence of fibres through the
material results in an overall anisotropic response of the tissue. The formulation of our model is
based on the balance of strain-energy potentials proposed by Simo et al., (1985); see also Simo et al.
(1991), Weiss et al. (1996), and Blemker et al. (2005). Our approach is to numerically approximate
the displacements $\u$, internal pressures $p$, and dilations $J$ of the tissues so that the total strain-energy
of the system $E_{\rm tot}$ reaches a local optimum. The total strain-energy of the system is given by:
\begin{align}
 E_{\rm tot}(\u,p,J) = U_{\rm int}(\u,p,J) - W_{\rm ext}(\u),\label{eq:totalenergy}
\end{align}
where $U_{\rm int}$ denotes the internal strain-energy potential of the muscle and $W_{\rm ext}$ denotes the external work on the system. In other words, we seek a state $(\u, p, J)$ such that the first variation of the total
strain-energy $DE_{\rm tot}$ is zero:
\begin{align}
 DE_{\rm tot}(\u,p,J) = 0.\label{eq:firstvariation}
\end{align}
To approximate the solutions $(\u, p, J)$ of \autoref{eq:firstvariation} we used the finite element method, and to approximate the integrals that are computed as a part of this method, we used the quadrature rule which involves quadrature points and weights. Therefore, $\u$, $p$, and $J$ are only known at the quadrature points. See \autoref{appendix:form} for more details on the formulation of our problem. The model was implemented in the finite element library deal.II version 8.5 (Arndt et al., 2017).

\subsubsection{Material Properties}\label{sect:materialproperties}
The fibres in the fibre-reinforced composite material represent the behaviour of the
myofilaments in muscle that develop both active (contractile element) and passive (parallel elastic)
forces, and the tissue surrounding the fibres that we refer to as base material, represents the behaviour
of the additional intra- and extracellular components that include connective tissue such as ECM,
blood, and other materials within whole muscle. We formulated the active and passive fibre curves as
trigonometric polynomial and second-order piecewise polynomial fits of experimental data (Winters
et al., 2011). These curves (Figure 1A) are similar in shape to the Bézier curves presented in Ross et
al., 2018a but are not parametric. To model the base material properties of the muscle, we used a
Yeoh model (Yeoh, 1993) fit to experimental data for tensile loading of muscle in the across-fibre
direction (Mohammadkhah et al., 2016) (Figure 1C). Because the properties of the fibres only act in
the along-fibre direction, the tensile across-fibre data from Mohammadkhah and colleagues (2016)
likely only represents the properties of the base material surrounding the fibres. We assumed that the
base material is isotropic and so contributes to the muscle tissue response in all directions.

While the block geometries in this paper are composed of only muscle and do not account for
the effects of aponeuroses, we included both a superficial and deep aponeurosis in the MRI-derived
geometries to better replicate the behaviour of the in vivo medial gastrocnemius. As with the muscle
tissue, we modelled the aponeurosis tissue as fibre-reinforced composite material. However, while
the fibres in the muscle tissue produce both active and passive forces, the fibres in the aponeurosis
tissue produce only passive forces and represent the behaviour of the bundles of collagen fibres
within the connective tissue. Given that tendon is an extension of aponeurosis and likely has similar
composition and collagen properties, we fit the passive fibre curve to experimental stress-stretch data
for tendon (Dick et al., 2016). This passive fibre curve is of the same form as the piecewise
polynomial that we used for the muscle fibre passive curve and can be seen in Figure 1B. To model
the base material properties of the aponeurosis, we fit the model from Yeoh (1993) to transverse
tensile loading data for aponeurosis (Azizi et al., 2009; Figure 1D).

We modelled both the muscle and aponeurosis tissue as nearly incompressible, with a
volumetric strain-energy potential describing the energetic cost of the compression that does occur in
the muscle. These volumetric strain-energy potentials were described by their bulk modulus $\kappa$, that
took values of $\kappa = 10^6$ Pa for the muscle and $\kappa = 10^8$ Pa for the aponeurosis. Finally, we set the
maximum isometric stress of the tissues to 200 kPa.

\subsection{Experimental data collection}\label{section:datacollection}
We collected surface geometry and internal architecture data for the medial gastrocnemius
(MG) muscle using magnetic resonance (MRI) and diffusion tensor (DTI) images of the lower limb.
These data were to provide initial geometries for model simulations of muscle contraction (from the
resting condition), and to provide deformed geometries and architectures during isometric contraction
to validate the simulation outputs from the finite element model of muscle contractions.

Four female participants (age 29 $\pm$ 4 years mean $\pm$ standard deviation.) with no recent history of
musculoskeletal disease or injury took part in this study. All procedures conformed to the Declaration
of Helsinki (2008) and were approved by University of New South Wales’ Human Research Ethics
Committee HREC (approval HC17106). We obtained informed consent from all individual
participants included in the study. Details of the MRI acquisition and data analysis can be found in
Section 7.2. Briefly, we had participants lie supine in an MRI scanner with their right knee slightly
flexed, their right foot strapped to a footplate and their ankle at 5 degrees plantarflexion. We instructed
participants to generate plantarflexion torques of 10 \% (twice) and 20 \% (once) of their maximum
voluntary plantarflexion torque while we imaged their right lower leg: we provided visual feedback
of the plantarflexion force to help participants maintain constant plantarflexion torque during the 2.5-
minute scans.

We calculated fibre orientations (primary eigenvectors of the diffusion tensor) both while the
muscle was relaxed, and during the active contractions where the participants generated constant
plantarflexion torque from the DTI scans, and we created 3D surface models of the medial
gastrocnemius from the anatomical MRI scans during these active and passive contractions (Figure
2).

\subsection{Model simulations}\label{section:simulations}
\subsubsection{Simulations of block geometries}\label{sections:blocks}
We constructed a series of blocks of parallel-fibred and unipennate muscle with cuboid geometries and no aponeurosis (Figure 3). We defined the length of the blocks as the distance between the positive and negative x-faces in the x-direction. The muscle fibres were parallel to each other and the xz plane, but oriented at an initial pennation angle $\beta_0$ away from the x-direction. We determined the cross-sectional area CSA of each muscle block from its initial configuration $V_0$ as the area of the
cross-section in the yz plane. The muscle blocks had faces in the positive and negative x, y, and z sides (for $V_0$ ) that deformed during contraction. The standard dimensions for the muscle blocks were 30x10x10 mm, however, we varied CSA and the block volumes $Vol$ by 15\%, and $\beta_0$ from 0 to 37 degrees so that the effects of $\beta_0$, CSA, and $Vol$ could be independently tested.

We simulated contractions of the muscle blocks using the FEM model. To hold the end of the blocks in an isometric state, we imposed kinematic constraints on the positive and negative x end faces in all three directions. We set the initial length of the fibres to their optimal length $(\lambda_{\rm iso} = 1)$ and linearly ramped the activation from 0\% to 100\% over 10 time steps. For these blocks containing only muscle tissue, the simulations would only converge to an activation of 100\% when $\beta$ was greater
than 5 degrees, so we increased the stiffness of the base material using a scaling factor s base of 1.5 to allow the model to converge to maximum activation when $\beta_0$ was 5 degrees or less.

\subsubsection{Simulations of MRI-derived geometries}\label{sections:mri}
We described the surface geometry of the MG using hexahedral meshes to bound the MRI
derived of the muscles at rest for all four participants. To do this, we outlined the shape of the muscle on all scan slices where the muscle was visible and then used these outlines to create a surface model of the muscle with 100 nodes. We converted the surface model to a volumetric tetrahedral mesh and then to a hexahedral mesh in GMSH format using GIBBON Toolbox and custom-built Matlab algorithms (Matlab 2018b; GIBBON Toolbox) to make it compatible with the finite element modelling software.

Large parts of the MG surface are covered by aponeurosis, so unlike the block simulations, we included superficial and deep aponeurosis in these simulations to better mimic the behaviour of the whole muscle during contraction. Aponeuroses are thin and difficult to discern on MRI scans so we identified them as regions where the muscle fibres intersect with the muscle surface: these fibres (4,039-7,745 per participant) were tracked using tractography methods on the DTI data, described in 
Bolsterlee et al. (2019). We added new hexahedral elements to the outside of the muscle surface where the aponeuroses had been identified. These elements tapered in thickness along the muscle’s length (2 mm thick) where they merged with the external tendon down to 1 mm thick at the other end (Figure 2) and were assigned aponeurosis properties for the model.

The DTI-derived muscle fibre orientations provided an opportunity to populate the MG-based geometry with the actual $\beta_0$ at each point. To achieve this, we sampled local fibre orientations at 2mm intervals along muscle fibres from DTI-derived fibre tracts, and then assigned fibre orientations to quadrature points of all muscle elements using nearest neighbor interpolation (or extrapolation for the most proximal part of the muscle for which no DTI data were available). We set the fibre orientations of quadrature points inside the aponeurosis to be tangential to the muscle surface with a zero y-component, i.e. parallel to the muscle’s surface and nearly parallel to the muscle’s long axis. We additionally evaluated simulations of the MG-based muscle using constant pennation angle $\beta_0$ through the muscle, to compare directly with results from the isolated muscle blocks.

We simulated fixed-end contractions of the MRI-derived geometries for the medial
gastrocnemius, up to 100\% muscle activation in increments of 10\%, by applying kinematic
constraints in all three orthogonal directions to the most proximal faces of the superficial aponeurosis
and the most distal faces of the deep aponeurosis.

\subsection{Post-processing and data analysis}\label{section:postprocessing}
The model geometries used for the FEM simulations were each bounded by their surface. For the block simulations, we characterised the faces of the blocks (-x, +x, -y, +y, -z and +z faces) for the undeformed state $V_0$, and then followed these for each deformed state $V$. The length of the muscle block $l$ is the distance between the -x and +x faces and was normalized $\hat{l}$to the length in the undeformed state. The strain $\varepsilon$ between the faces is the change in distance between opposite faces,
normalized to their initial separation in the undeformed state. The geometries from the medial gastrocnemius muscles from the MRI scans had no distinct faces and so we characterized the changes in width and depth from the whole surface. We sampled cross-sections of the surfaces at 10\% intervals along the muscle length: the width was the maximum width of the section, and the depth was given by the cross-sectional area of that section divided by its width. 

We defined muscle bulging as displacement of the muscle’s surface in the direction perpendicular to the surface. We calculated bulging using distance maps. A distance map is a 3D regular grid of points in which the absolute value of each grid point equals the distance to the nearest point on the surface model. To determine muscle bulging during contraction, a distance map of the muscle surface at rest $V_0$ was created. We aligned the surface for the current state $V$ with that for the undeformed state $V_0$
using principal component analysis. The distance map was then interpolated at the nodes of the aligned current state. The value associated with each node thus approximates the distance to the nearest point of the muscle at rest, allowing for quantification of muscle bulging patterns: the sign indicates whether a point is inside (negative) or outside (positive) the muscle surface. Muscle bulging was calculated using this same approach for both the MRI geometries, and the FEM simulations of the medial gastrocnemius.

The FEM model calculates tissue properties across a set of quadrature points within each model: 128,000 quadrature points for the muscle blocks, and approximately 37,000 for the medial gastrocnemius geometries. We defined an orientation and stretch (normalized length) at each quadrature point. The pennation angle in the undeformed $\beta_0$ and current $\beta$ states were calculated as the angle between the fibre orientations and the x-axis: this is an angle in 3D space, similar to the 3D
pennation angles defined by Rana et al. (2013). The fibre stretch $\lambda_{\rm tot}$ gives the normalized length of the tissue in the direction of the fibres at each quadrature points. These pennation angles $\beta$ and fibre stretches $\lambda_{\rm tot}$ are thus calculated for local regions within the muscle tissue, and so we sometimes reported them as their mean value across the whole tissue or block.

We calculated forces $F$ as the magnitude of force perpendicular to a face or plane within the muscle, and the stress $\sigma$ as that force divided by the area of that face or plane in the current state of the simulation. The strain-energies are initially calculated as strain energy-densities $\Psi$, which are the strain-energy for a given volume of tissue, in units Jm$^{-3}$. The FEM calculates $\Psi$ for every quadrature point, and so we calculated the overall strain energy-density from the weighted mean of $\Psi$ where it is weighted by the local dilation at each quadrature point. The strain-energy potential $U$ is the strain-energy in the tissue, in units of Joules. We calculated $U$ by integrating $\Psi$ across the volume of muscle tissue.

We identified a 30x10x10 mm region in the centre of the MG simulation, and
compared its results to simulated values for an isolated block of muscle tissue, taking both blocks with $\beta_0$ = 25 degress, and $s_{\rm base} = 1$. Symbols used to reference the post-processing parameters are shown in Table 1.

\section{Results}\label{section:results}
\subsection{Simulations of block geometries}\label{section:blocksimulations}
Isometric constraints were in the same direction as the fibre orientation for the parallel fibred
($\beta_0 =0$ degrees) blocks, and so their fibres showed no net shortening. Instead, the volume of the blocks
showed a marginal increase during activation, with the fibre stretch $\bar{\lambda}_{\rm iso}$ increasing minimally (Figure
4A). The mean pennation angle for the parallel fibred block was $\bar{\beta}= 0$ degrees at full activation. On the
other hand, the isometric constraints were not in the same direction as the fibre orientation for the
pennate ($\beta_0 >0$ degrees) blocks, and so their fibres underwent a net shortening during activation (Figure 4D).
Indeed, at 100\% activation the $\beta_0 =30$ degrees block shortened to $\bar{\lambda}_{\rm iso} = 0.86$, and its pennation angle
increased to 33.6 degrees Figure 4D-E).

Stresses normal to the mean fibre direction, through the centre of the muscle blocks, increased
as activation increased and are shown for the fully active conditions (Figure 5A). These stresses had
components due to the different strain-energy potentials. The stresses due to the active-fibre and the
volumetric strain-energy potentials both acted to shorten the fibres, whilst the stress from the base
material acted to resist shortening. For the parallel-fibred case ($\beta_0 =0$ degrees), the stress from the volumetric
component was a large proportion of the total stress, and there was little resisting stress from the base
material. These features transitioned as the pennation angle increased, and the $\beta_0 =30$ degrees block had the
least stress from the active fibre strain-energy potential, and the greatest resistive stress from the base
material strain-energy potential. As the pennation angle increased, the normal stress to the fibres had
a smaller component in the line-of-action (x-direction) of the blocks. Indeed, the parallel fibred block
($\beta_0 =0$ degrees) developed a force of $F_x =19.06$ N in its line of action, whereas the pennate block ($\beta_0 =30$ degrees)
developed a reduced force of 10.70 N (Figure 4C, F).

The x-stress on the x-face increased as activation increased (Fig. 6A, D). However, the y-
stress on the y-face and the z-stress on the z-face were minimal, due to these faces being
unconstrained. Nonetheless, stresses in the y- and z-directions developed within the blocks of muscle
when the muscle activated. Within the blocks, stresses in the y- and z-directions were transversely
isotropic for the $\beta_0 =0$ degrees block, but showed increasing asymmetry as the pennation angle increased. In
general the y- stress was larger than the z-stress, and both acted to expand the muscle block, however,
at larger pennation angles ($\beta_0 >25$ degrees) the z-stress became minimal or compressive. The muscle blocks
deformed in 3D. For both the parallel and pennate example, the x-faces remained isometric, and so no
x-strain was recorded. For the parallel-fibred block, the small increase in volume resulted in a small,
but isotropic, strain in the y- and z-directions (Figure 6B). There was a transition pennation angle at
$\beta_0\approx15$ degrees below which the z-strain was positive with the z-faces increasing in separation, and above
which the z-strain was negative with the z-faces becoming closer during activation (Figure 7). Small
changes in the active-fibre strain-energy potential in the parallel-fibred block were largely balanced
by increases in the volumetric strain-energy potential: here the changes in passive-fibre and base
material strain-energy potentials were much smaller (Figure 8C). By contrast, the active-fibre strain-
energy potential showed a larger change in the pennate block of muscle that, in this case, was largely
balanced by increases in the base material strain-energy potential: here the changes in volumetric and
passive-fibre strain-energy potentials were much smaller (Figure 6D).

When the parallel-fibred muscle block ($\beta_0 =0$ degrees) was stretched or shortened to different lengths
before the activation began, the balance of the strain-energy potentials changed within the muscle.
When the muscle block was fully active, the component of the stress due to the active-fibre strain-
energy potential acted to shorten the muscle at all muscle lengths tested. The components of stress
due to the volumetric and base material strain-energy potentials both acted to resist shortening at the
short muscle lengths ($\hat{l}< 0.9$), and thus contributed to a reduction to the force in the line of action $F_x$
(Figure 5B). At the longer muscle lengths, the components of stress due to the volumetric, base material and passive-fibre strain-energy potentials all acted to resist lengthening ($\hat{l}< 1.1$).
Interestingly, the contribution of the passive-fibre to the overall resistive force was less than that for
the base material and also the volumetric components (Figure 5B).

The components of the strain energy-density showed little change with cross-sectional area of
the muscle blocks, but a pronounced change with pennation angle (Figure 8). There were very few
points in the muscle blocks that showed an increase in fibre stretch at full activation, and so the strain
energy-density for the passive-fibre component was small for these simulations (Figure 8B).
However, the strain energy-density for the base material increased in an almost linear fashion with
pennation angle ($r^2 =0.99$, Figure 8D). The strain-energy potential from the base material acted to
resist the fibre shortening, and the strain-energy potential from the volumetric and active-fibre
components acted to shorten the fibres (Figure 5A). The strain energy-density for the active fibres
increased in magnitude at greater pennation angles (Figure 8A), whereas the volumetric component
of the strain energy-density decreased at higher pennation angles (Figure 8C). The stress in the line of
action of the muscle blocks (x-stress on x-face) remained high for pennation angles up to 15-20 degrees
(Figure 8E) and showed substantial reduction for pennation angles greater than 20 degrees Figure 8E).

The isolated block of muscle showed similar deformations and strain-energy densities as to
the block of similar size extracted from the simulation in the MRI-derived MG geometry (with
$\beta_0 =25$ degrees for both; Figure 9B). There was a greater spread of values in the isolated block, due to the
proximity of the isometric constraints on the faces, however, the median fibre strain, dilation, and
pennation angle were different by less than 1\% or 1 degrees for these simulations. Additionally, the strain
energy densities had a close match for the two conditions (Figure 9C).

\subsection{Simulations of MRI-derived geometries}\label{section:mrisimulations}
The simulations and the DTI data both showed increases in pennation angle $\beta$ during contractions
(Figure 10). However, this increase was larger for the DTI data (11 degrees at 20\% contraction) than the
simulations (3 degrees). The simulations and the MRI data both showed relatively small changes ($< 2\%$) in
muscle width and depth at 10\% activation. At the most distal end, the model decreased its depth
slightly whilst the depth increased in the most proximal regions (Figure 11 A-B). Changes in width
and depth were larger and more heterogeneous along the muscle’s length (Figure 11C-D) at 20\%
activation. The proximal region increased in width and decreased in depth while the distal part
decreased in width and increased in depth. On average, changes in width were similar between
simulations and MRI measurements. However, the simulation with DTI-derived fibre orientations did
not predict the decrease in depth observed in MRI data for 20\% activation. Adjusting the initial
pennation angle of the model to $\beta_0=25$ degrees resulted in a closer match between DTI-derived and
simulated fibre orientations at 20\% activation (Figure 10), and a close match in magnitude and
pattern of muscle depth change between MRI and simulations (Figure 11C-D). The adjusted model
also resulted in a close match of 3D muscle bulging patterns predicted by the model and as measured
from MRI (Figure 12).

\section{Discussion}
This study investigates the energetic mechanisms within muscle tissue during isometric
contractions. The pennate blocks of muscle ($\beta_0  >0$ degrees) that we modelled showed general features of
contraction that have been typically reported in both animal and human studies (Héroux et al., 2016;
Kawakami et al. 1998). The fibres shortened ($\lambda_{\rm tot} <1$) and rotated to greater pennation angles during
contraction, even though the ends of the blocks were fixed (Figure 4). An asymmetry developed to
the stress in the transverse (yz) plane when the muscle was active. Changes to the tissue shape were
governed by the isotropic base material properties and the volumetric strain energies: because there
was an asymmetry to the stress across the muscle, this resulted in an asymmetry to the transverse
tissue deformation that was dependent on pennation angle (Figure 7). These findings explain a
mechanism that can result in transverse anisotropy within a muscle, that we have previously reported
(Randhawa and Wakeling, 2018).

Interestingly, the volume of the muscle blocks increased during contraction to a small extent
(0.6-0.9\% for the 20\% activation MG simulations; Figure 9B). The mechanism for this increase is
described in section 5.1. The extent of the increase in volume is related to the choice of the bulk
modulus $\kappa$ of the tissue that is used to calculate the volumetric strain-energy potential. However,
previous studies have shown that $\kappa$ can be varied across a wide range of magnitudes and still result in
similar predictions of tissue deformation (Gardiner and Weiss, 2001), and here we used a value
consistent with previous our studies (Rahemi et al., 2014, 2015). Our finding that muscle volume can
change is consistent with a number of previous studies investigating muscle at different scales
(Neering et al., 1991; Smith et al., 2011; Bolsterlee et al., 2017). Intriguingly, the volume of muscle
tissue will tend to increase with the muscle bulging transversely, even for an isometric contraction for
blocks with zero pennation angle. However, this is consistent with the finding that regions of single
fibres can increase in volume during isometric contraction (Neering et al., 1991). Local increases in
volume had previously been explained due to cytoskeletal effects (Neering et al., 1991); in our
simulations the cytoskeleton is represented as part of the base material, and we show that as energy is
redistributed to the base material and volumetric components, there is a tendency for the volume to
increase. The changes in volume were not uniform through the blocks of muscle. Indeed, variations
in bulging along a muscle belly have also been reported in both human and rabbit muscle (Raiteri et
al., 2016; Böl et al., 2013). It should be noted that increases in intramuscular pressure during
contraction may expel blood from the muscle (Barnes, 1986; Sjøgaard et al., 1988), acting to
decrease the whole muscle volume; this may occlude local increases in volume of the muscle tissue
due to the volumetric strain-energy potential.

Evaluating the muscle model within the actual MRI-derived geometry of the medial
gastrocnemius allowed us to qualitatively validate the outputs from the model. When muscle
contracts it develops force and changes length (or more exactly changes shape in 3D). Direct
measures of muscle force are virtually impossible to make in humans, and even in the few animal
studies where they are measured, the forces would typically only be measured in one-dimension.
Thus, complete force and deformation data sets for validating 3D muscle models are sparse for
animal studies (Böl et al., 2013), and non-existent for human studies. However, 3D muscle models
have previously been validated against deformations of contracting muscle for both animal (Tang et
al., 2007) and human (Blemker et al., 2005; Böl et al., 2011) studies. MRI allows 3D deformations of
the whole muscle geometry to be measured, allowing for validation of the surface deformations that
were generated by the muscle model. It should be noted that the MRI images of the medial
gastrocnemius were from the intact leg, and thus subject to external forces and boundary constraints
(from surrounding tissues) that were not replicated in the model here. Additionally, the MRI images were for muscle contractions with fixed joint angles, however, due to stretch in the tendons the
muscle belly would undergo some shortening (approximately 3 mm during 20\% contractions as
measured from the MRI scans), and thus the model constraints should not be considered as an exact
match of the MRI experimental situation. Nonetheless, there was a close match of the deformations
of the surface geometry between the MRI and FEM model results (Figure 12). Additionally, a block
of muscle identified within the actual MRI-derived geometry showed similar patterns of strain-
energy densities as for an isolated block of muscle (Figure 9C). These results give confidence that the
mechanisms of contraction identified for the blocks of muscle tissue are realistic.

Novel results from this study are that regions of the muscle are displaced inwards, particularly
under the aponeuroses, whereas other regions are displaced outwards, predominantly at the ends and
edges of the muscle, and this bulging is apparent in both the MRI images and the FEM model results.
Previous ultrasound studies have suggested that tissue deformations and volume changes predicted
from imaging the middle region of the muscle belly may not represent deformations along the entire
muscle if deformations and volume changes vary along its length (Raiteri et al., 2016; Randhawa et
al., 2018), and these suggestions are now supported by the results from this study.

\subsubsection{Strain-energy distribution through contracting muscle}\label{section:energydistribution}
When the muscle contracts it increases in its free energy, with this energy being derived from
the hydrolysis of ATP to ADP within the muscle fibres (Woledge et al., 1985; Aidley, 1998). There
is only finite free energy available from hydrolysis of ATP within the muscle, governed by the
availability of nutrients and ATP, therefore, there is a limit to the work that can be done during a
muscle contraction. The mechanical work that can be done by a contracting sarcomere in its line-of-
action is an intrinsic property of the sarcomere, is given by the area under the active force-length
curve, and has an energy-density of approximately $1.5 \cdot 10^5$ Jm$^-3$ (Weis-Fogh and Alexander, 1978).
Strain-energy potentials develop in the fibres of our FEM model during active contraction: these
fibres represent the contractile elements within the myofilaments in muscle. Within the
myofilaments, the cross-bridges contain energy when they attach between the actin and myosin as
part of the cross-bridge cycle (Williams et al., 2010) like a set of taught springs. This energy is
partially redistributed to the thick and thin myofilaments (Williams et al., 2012), with additional
energy being released during the power stroke of the cross-bridge cycle. Strain-energy potentials are
also redistributed to the titin filaments that are large proteins that span from the M-line to the Z-disc
(Gregorio et al., 1999) and likely account for the majority of the passive-fibre strain energy. Base
material strain-energy potential can develop in the bulk muscle tissue within the muscle fibres
(excluding the myofilament fraction), connective tissue surrounding the muscle fibres such as the
extracellular matrix, and in sheets of connective tissue that form the aponeuroses, internal and
external tendons. Energy is also used to change the muscle volume. Whilst muscle is often assumed
to be incompressible, small changes in volume can occur in fibres (Neering et al., 1991), bundles of
fibres called fascicles (Smith et al., 2011) and in the whole muscle (Bolsterlee et al., 2017): these
changes in volume are energy-consuming processes. Additionally, energy is required for the
acceleration of the tissue mass within the muscle to overcome its inertia during rapid movements
(Ross et al., 2018, Ross et al., 2020). Finally, energy is transduced to mechanical work at the surface
of the muscle where the muscle changes shape and exerts forces on surrounding structures (Siebert et
al., 2012; Ryan et al., 2020).

Muscle force developed in the line-of-action is given by the x-component of force on the
positive and negative x-faces from the blocks of muscle in this study. As the strain-energy potentials
within the muscle are redistributed between the different components of energy (volumetric, base
material and active- and passive-fibre strain-energies), and because the strain-energy potentials in
both the base material and volumetric components are distributed across all three dimensions, the
force that can be developed in the line-of-action will be less than that could be generated by just the
contractile elements alone. This is a fundamental consequence of encasing the model fibres
(representing the muscles’ contractile elements) within the bulk muscle tissue. In addition, the
muscle fibres develop non-uniform stretches throughout the muscle (Figure 9B), and so the muscle is
further unable to contract with all its fibres at their optimal length, and thus the whole muscle tissue
will always contract at forces less than the theoretical maximum isometric force. Even when the
stresses and forces are considered relative to the fibre orientation, we find that redistribution of
strain-energy potentials through the muscle tissues results in contractile stresses (normal to the fibre
direction: Figure 5A) being developed by both volumetric and active-fibre components when the
fibre stretch $\lambda_{\rm tot} < 1$, and additionally from passive-fibre and base material components at longer
muscle ($\hat{l}> 1$) and fibre lengths (Figure 5B).

Our computational results suggest that muscle ($\beta_0  =0$ degrees) bulges slightly due to its base material
properties, even when it is isometric and there is no series elasticity such as tendon that could allow
the muscle belly to shorten. This again can be explained in terms of the energy redistribution. The
free energy in the muscle increases during the activation process and will be redistributed across
fibre, base material and volumetric strain-energy potentials. The energetically favourable state
identified in our simulations occurred with a small increase in tissue volume, due to transverse
expansion of the fibres (in the yz-plane). Thus, muscle bulging should not only be considered to be a
consequence of muscle shortening leading to an increase in cross-section to maintain a constant
volume (eg. Azizi et al., 2008; Siebert et al., 2012; Azizi et al., 2017), but may also occur due to the
biological tissues showing small changes in volume, even for isometric contractions. This mechanism
is consistent with the finding that even single fibres can bulge during isometric contraction (Neering
et al., 1991).

The stresses in the tissues are defined as the first variation of the strain energy-densities
(\autoref{eq:firstvariation}). We need to integrate these stresses in order to obtain the strain-energy potentials from
known values of the stress. However, this would only provide a change of the energy; the integral of
the stress equals the difference of the strain-energy potentials between two different states. The
strain-energy potentials presented in this study are relative to the undeformed state $V_0$ of the whole
muscle.

This paper focusses on the internal energy within the muscle during isometric contraction.
However, it should be noted that the whole energy balance will also include external work done at
the surface of the muscle (\autoref{eq:totalenergy} in \autoref{appendix:form}). For the case of the isometric block simulations
at the initial muscle length, this external work is zero. However, we had to apply external work to the
x-faces of the system to lengthen of shorten the muscle blocks for Figure 5B. It should be noted that
external work could be done at any point on the muscle surface, for instance transverse compression
of the muscle (in the yz-plane), and this is the topic of our companion paper (Ryan et al., 2020).
Additionally, kinematic energy is required to accelerate the tissue mass, and should be included to
the energy balance (\autoref{eq:stressfibre} in \autoref{appendix:form}) to understand the effect of muscle mass on dynamic
contractions of whole muscle: this is the topic of our second companion paper (Ross et al., 2020).

\subsubsection{Implications for muscle structure and function}\label{section:musclefunction}
Here we show that considerable strain-energy potential develops in the base material during
these isometric muscle contractions, with this strain-energy potential increasing as muscles become
more pennate (Figure 6F). The base material resists the contractile force in the line of action, and so
the increasing involvement of the base material results in a progressive suppression of the muscle
force for more pennate muscle. In this study we have implemented the base material as an isotropic
material. However, elements of the base material do have anisotropy that is conferred by their
structure such as the network of connected tunnels forming the endomysium that contain a feltwork
of collagen fibres (Purslow and Trotter, 1994). Lumped constitutive models, such as used here and by
eg. Blemker et al. (2005) and Rahemi et al. (2015), are unlikely to capture the details of base material
anisotropy and the asymmetric response to compression and tension (Böl et al. 2012; Gindre et al.
2013). Anisotropy in the base material properties is most pronounced when the tissue is in tension
and may be reasonably disregarded for compressive tests with $\lambda_{\rm tot} <1$ (Böl et al., 2012), as is the case
for all the block tests in Figures 4, 5A 6 \& 8. Nevertheless, we suggest here that for all the isometric
tests in this study (regardless of the degree of anisotropy in the base material), the base material
would still act to resist muscle deformation and the force in the line-of-action; however, the extent to
which the base material interacts with the fibre direction, and thus muscle pennation, depends on the
extent of its anisotropy. We additionally show that even though the changes in volumetric strain-
energy potential are small, relative to the base material strain-energy potential, the contribution of the
volumetric strain-energy potential to the contractile stress and force can be considerable (Figure 5).
Thus, even though the contribution of volumetric and base material strain-energy potentials has been
largely ignored to date in considerations of whole muscle force and deformation, we suggest that they
play an important and significant role in the 3D structure and function of whole contractions.
Subsequently, this finding and study highlight how little we currently know about these processes,
and how important it will be to further characterise and implement the base material and volumetric
properties of muscle as we continue to learn about 3D function of whole muscle contractions.

The muscle force and stress in the line of action was reduced at the higher pennation angles
(Figure 5A, 8E). This may be partly due to a region of muscle tissue in the middle of the blocks
having fibres that did not connect directly to the x-faces of the block. These ‘unsupported’ fibres
would still develop strain-energy potentials as they deformed their base material during shortening;
however, it is possible that lateral transmission of force across the base material was not fully
accommodated by the model parameters used, thus these fibres may not have fully contributed to the
$F_x$ forces experienced by the x-faces of the block. Nonetheless, our results show no evidence that
increased pennation angle $\beta_0$  causes an increase in the force in the line of action of the muscle.
Instead, the results from this study support the notion that the functional benefit of pennation in
muscle may be to reduce the metabolic cost of contraction (Biewener, 2003), or allowing the fibres to
reduce their contractile velocity and thus be better geared for dynamic force production (Azizi et al.,
2008), rather than to increase the muscle force for isometric contractions per se (Alexander, 1983;
Lieber and Fridén, 2000; Biewener, 2003).

We have previously shown how intramuscular fat decreases the force and stress that can be
produced by isometrically contracting muscle (Rahemi et al., 2015), using a similar FEM approach to
this current paper. In the fat study (Rahemi et al., 2015) we implemented the intramuscular fat into
model simulations in a number of ways and found that all the fatty models generated lower fibre
stress and muscle force than their lean counterparts. This effect was due to the higher base-material
stiffness of the tissue in the fatty models. This fat study highlighted how the material properties of the
base material may cause profound and important changes to muscle contractile performance, and this was due to the same mechanisms of energy redistribution as we now describe in this current study.
There are a range of muscle conditions and impairments that have fibrotic tissue, changing their
stiffness, and this energetic framework now provides an approach for us to understand how such
conditions lead to loss of muscle function. For instance, altered material properties of muscle tissue
post stroke (Lee et al., 2015) and with cerebral palsy (Lieber \& Fridén, 2019) have been linked to
increases in collagenous connective tissue within the muscle (Lieber and Ward, 2013). Whilst it is
possible to measure proxies of muscle tissue stiffness with shear wave ultrasound elastography (Lee
et al., 2015), it is difficult to partition these changes between the passive stiffness of the fibres, or the
stiffness of the base material. Nonetheless, increased collagen content in the extracellular matrix (that
contributes to the base material properties in this study) causes an increase in the passive stiffness of
the muscle in mice (Meyer and Lieber, 2011; Wood et al., 2014). As such, we suggest that
understanding how altered tissue properties affect the energetic consequences of muscle deformations
will allow us to understand muscle impairments in greater detail.

\section{Conclusions}
\begin{enumerate}
 \item Strain-energy potentials develop within muscle tissue during contraction, even for isometric contractions where there is no external work.
\item Strain-energy potentials are distributed across different components within the muscle: the
contractile elements as the active- and passive-fibre strain-energy potentials, the cellular and
extracellular components as the base material strain-energy potentials and the volumetric component
to enforce the nearly isovolumetric constraints. The balance of this strain-energy distribution may
seem counter-intuitive, and it depends on the length of the muscle and the orientation of its fibres.
\item The volumetric and base material strain-energy potentials redistribute the energy into all three
dimensions and affect the 3D deformations of the muscle.
\item Strain-energy potentials taken up by the base material and sometimes the volumetric (for shorter
muscle lengths) components result in less strain-energy available for force development in the line-
of-action of the muscle.
\item The muscle force in the line of action is thus never as great as could be predicted from the intrinsic
properties of the contractile elements alone. This loss in force gets more pronounced for highly
pennate muscle, particularly where $\beta_0 > 20$ degrees.
\end{enumerate}

\appendix
\section{Details of model formulation}\label{appendix:form}
See Section 3.1.1. for the general formulation of the finite element model and Table 2 for the definitions and notation of the variables that we used in this section.

The Cauchy stress in the tissues $\bsigma$ is defined as the first variation of the internal strain energy-density $\Psi_{\rm int}$; see \autoref{eq:definitionstress}. To obtain the strain-energy potentials, we integrate the Cauchy stress. However, this provides a change of the energy; the integral of the stress equals the difference of the strain-energy potential between two different states; see \autoref{eq:definitionenergies}. The strain-energy potentials reported in this study are relative to the undeformed state $V_0$ of the whole muscle. In our formulation, the undeformed configuration has a zero strain-energy potential as all internal stresses and external
forces in the system are zero.

This paper focuses on the change of the internal strain-energy potential $U_int$ within the muscle
during isometric contraction. However, it should be noted that the whole energy balance will also
include external work done at the surface of the muscle (\autoref{eq:totalenergy}). For the case of the isometric
block simulations at the initial muscle length, this external work is zero. However, when the muscle
block was initially stretched or shortened to a new passive length, with traction being applied to the
positive x-face, external work is added to the system.

We assume that a muscle occupies an initial volume $V_0$ and denote its surface by $S_0$. Using the principles of continuum mechanics, a point $q_0$ in $V_0$ can be tracked over time after the muscle in $V_0$ has seen some deformation. Consider $V$ as the volume that the muscle in $V_0$ has been deformed at time $t$. In continuum mechanics we assume that there exists a unique point $q$ in $V$ such that $q = q(q_0,t)$. This means that the point $q$ is the representation of the point $q_0$ in the current configuration of the muscle. The displacement vector at the point $q_0$ in $V_0$ is defined as the vector formed by the points $q$ and $q_0$, that is
\begin{align*}
    \u(q_0,t) := q(q_0,t) - q_0.
\end{align*}
The deformation of the tissues is defined as the gradient of this displacement. This deformation is quantified as a tensor of infinitesimal changes of points $q_0$ in $V_0$ at time $t$. Mathematically, this is defined as:
\begin{align*}
    \ff:=\ii + \nabla_0\u,
\end{align*}
where $\ii$ is the 3-by-3 identity matrix and $\nabla_0$ is the gradient of a vector with respect to the initial configuration $V_0$. We denote
the dilation of the tissues by $J$, and the internal pressure in the tissues by $p$.

The deformation tensor F provides a change of variables for integrals over $V_0$ and $V$ as well as
over the surfaces $S_0$ and $S$. The following change of variables hold:
\begin{align*}
    dV = I_3(\ff)\, dV_0,\quad\hat{\n}dS = I_3(\ff)\ff^{-\transpose}\hat\n_0\,dS_0,
\end{align*}
where $\hat{\n}$ is the unit normal vector on $S$ and $\hat{\n}_0$ is the unit normal vector on $S_0$, and $\ff^{-\transpose}$ is the transpose
tensor of the inverse tensor of the deformation tensor, $\ff^{-1}$. These changes of variables are used to compute volume and area of the deformed configurations of the muscle. We compute these as follows:
\begin{align*}
    Vol := \int_{V_0} I_3(\ff)\, dV_0,\quad A:= \int_{S_0}I_3(\ff) \sqrt{\hat{\n}_0^\transpose\ff^{-1}\ff^{-\transpose}\hat{\n}_0}\,dS_0.
\end{align*}
The fourth invariant $I_4$ of $\bb_{\rm iso}:= I_3(\bb)^{-1/3}\,\bb$, where $\bb$ is the left Cauchy tensor defined as $\bb := \ff^\transpose\,\ff$, is used in our model to represent the isovolumetric stretch of the fibres at the point $q_0$ in $V_0$, denoted by $\lambda_{\rm iso}$. Following the ideas in Simo et. al 1985, we consider normalized vectors $\hat{\a}_0 = \hat{\a}_0(q_0)$ that are tangential to the fibres at the point $q_0$ in $V_0$. We refer to $\hat{\a}_0$ as the initial orientation of the fibre at the point $q_0$ in $V_0$. The total stretch of the fibres at $q_0$ in $V_0$ is denoted by $\lambda_{\rm tot}$ and is defined as
\begin{align*}
    \lambda_{\rm tot} := I_3(\ff)^{1/3}\lambda_{\rm iso},\quad \lambda_{\rm iso} = I_4(\hat{\a}_0,\bb_{\rm iso}).
\end{align*}
Important elasticity tensors can be obtained from the internal strain-energy-density $\Psi_{\rm int}$. The Kirchhoff tensor $\btau$ and the Cauchy tensor $\bsigma$ are defined as
\begin{align}
    \btau := 2\,\bb\,\frac{\partial \Psi_{\rm int}}{\partial\bb},\quad \bsigma := I_3(\ff)^{-1}\,\btau.\label{eq:definitionstress}
\end{align}
As for the internal strain-energy potential, the strain-energy-density $\Psi_{\rm int}$ is split into a volume and an isovolumetric strain-energy-densities $\Psi_{\rm vol}$ and $\Psi_{\rm iso}$. Following Pelteret et. al 2012, the volume strain-energy-density $\Psi_{\rm vol}$ is defined as 
\begin{align*}
    \Psi_{\rm vol}(\u,p,J) := \frac{\kappa}{4}(J^2 - 2\,\log(J) - 1) + p\,(J-I_3(\ff)),
\end{align*}
where $\kappa > 0$ is the bulk modulus of the tissue. The value of the bulk modulus is different for muscle and aponeurosis, $\kappa_{\rm mus} = 10^6$ and $\kappa_{\rm apo} = 10^8$; see Rahemi et al. 2014. The pressure $p$ is added as a Lagrange multiplier in our system to ensure an accurate computation of the dilation $J$ in the tissues. 
 
The isovolumetric strain-energy-density is the sum of the base material and fibre strain-energy-densities,
\begin{align}
    \Psi_{\rm iso}(\u) := \Psi_{\rm base}(I_1(\bb_{\rm iso})) + \Psi_{\rm fibre}(\lambda_{\rm iso}).
\end{align}
We fit parameters from the models proposed by Yeoh, 1993 to data reported in Mohammadkhah et al., 2016 and Azizi et al., 2009 to obtain the mechanical properties for the base
material in both muscle and aponeurosis, respectively (see Section 3.1.2). The strain-energy-density $\Psi_{\rm base}$ and the scalar stress $\sigma_{\rm base}$ satisfy the following
\begin{align*}
    \frac{\partial\Psi_{\rm base}}{\partial I_1} = \sigma_{\rm base}(I_1),\quad I_1 = I_1(\bb_{\rm iso}).
\end{align*}
The muscle base material properties differ widely from those of the aponeurosis. We use $\Psi_{\rm mus,base}$ and $\sigma_{\rm mus, base}$ for muscle tissues, and use $\Psi_{\rm apo,base}$ and $\sigma_{\rm apo, base}$ for aponeurosis tissues.

Following Blemker et. al 2005, the scalar stress in the fibres $\sigma_{\rm fibre}$ is related to the strain-energy-density $\Psi_{\rm fibre}$ as follows
\begin{align}
    \lambda_{\rm iso}\frac{\partial \Psi_{\rm fibre}}{\partial\lambda_{\rm iso}} = \sigma_{\rm fibre}(\lambda_{\rm iso}).\label{eq:stressfibre}
\end{align}
For muscle tissues, passive and active properties are part of the normalized stress $\hat\sigma_{\rm fibre}$. We also consider the activation level of the tissues, denoted by $\hat a = \hat a(\x_0,t)$. The activation level only affects the active stresses. The stress due to fibre stretch in muscle fibres is defined as
\begin{align*}
    \hat\sigma_{mus,fibre}(\lambda_{\rm iso}) := \hat a(\x_0,t)\hat\sigma_{act}(\lambda_{\rm iso}) + \hat\sigma_{pas}(\lambda_{\rm iso}),
\end{align*}
where $\hat\sigma_{\rm act}$ represents the stress in the tissues due to active stretch of the muscle fibres and $\hat\sigma_{\rm pas}$ is the stress in the tissues due to passive stretch of the muscle fibres. Altogether, we can relate the stresses in the muscle fibres to the strain-energy-density by using \eqref{eq:stressfibre},
\begin{align*}
    \lambda_{\rm iso}\frac{\partial \Psi_{\rm mus, fibre}}{\partial\lambda_{\rm iso}} = \sigma_0\,\hat\sigma_{mus,fibre}(\lambda_{\rm iso}),
\end{align*}
where $\sigma_0$ is the maximum isometric stress of the contractile elements.

For aponeuroses, the stresses in the fibres are only due to passive stretch of the fibres. The strain-energy-density $\Psi_{\rm apo,fibre}$ and the scalar stress in the aponeurosis fibres $\sigma_{\rm apo,fibre}$ are defined exactly as in \eqref{eq:stressfibre},
\begin{align*}
    \lambda_{\rm iso}\frac{\partial \Psi_{\rm apo,fibre}}{\partial\lambda_{\rm iso}} = \sigma_{\rm apo,fibre}(\lambda_{\rm iso}).
\end{align*}
Our approach seeks a displacement $\u$, an internal pressure $p$ and a dilation $J$ which minimize the energy of the system. Let $E_{\rm tot}$ be the total strain-energy in the muscle. Under the assumption of a {\it quasi-static regime}, the total energy can be considered as the sum of the internal strain-energy potential and the work done on the system by external forces. Given an applied traction $\t$ on a part of the surface $S$ of $V$, denoted by $S_{\rm t}$, the total strain-energy potential of a deformed muscle in $V$ can be defined as
\begin{align}
    E_{\rm tot}(\u,p,J) = U_{\rm int}(\u,p,J) - W_{\rm ext}(\u),\label{eq:totalenergy}
\end{align}
where 
\begin{align}
    U_{\rm int}(\u,p,J) := \int_V\Psi_{\rm int}(\u,p,J)\,dV,\quad
    W_{\rm ext}(\u) := \int_{S_{\rm t}}\t\cdot\u\,dS_{\rm t}.\label{eq:definitionenergies}
\end{align}
We utilize the finite element method to approximate states $(\u,p,J)$ at which the first variation $DE_{\rm tot}$ of the total strain-energy potential is zero, that is we numerically solve the following equation for the unknowns $(\u,p,J)$:
\begin{align*}
    DE_{\rm tot}(\u,p,J) = 0. 
\end{align*}
The first variation of the total strain-energy gives the following set of equations to solve:
\begin{align*}
    \bdiv\,\bsigma(\u) = \zero,\,\,J - I_3(\u) = 0,\,\, p - \frac{\kappa}{2}(J-\frac{1}{J}) = 0\,\,\text{in}\,\,V,\\ \bsigma(\u)\hat\n = \t\,\,\text{on}\,\,S_{\rm t},\,\, \u = \zero\,\,\text{on}\,\,S_{\rm d},
\end{align*}
where the vector $\hat\n$ denotes the unit normal vector on $S$, and $S_{\rm d}$ stands for the part of $S$ on which the displacement $\u$ is prescribed. Note that in our model we clamp the displacement $\u$ on $S_{\rm d}$.

The change in the internal strain-energy potential reported in this paper is computed using the following definition
\begin{align}
    U_{\rm int}(\u,p,J) :=& \int_{V_0}\btau:\nabla\,\u\,dV_0 + U_{\rm int}(\u_0,p_0,J_0),\label{eq:changeinenergy}
\end{align}
where $(\u_0,p_0,J_0)$ is a known state of the displacement, pressure and dilation. We compute the
volumetric, isovolumetric, muscle base material, aponeurosis base material, muscle active-fibre,
muscle passive-fibre and aponeurosis fibre strain-energy potentials $U_{\rm vol}$, $U_{\rm iso}$, $U_{\rm mus,base}$, $U_{\rm act}$, $U_{\rm pas}$ and $U_{\rm apo,fibre}$ respectively are computed by using the formula given in \eqref{eq:changeinenergy},
\begin{align*}
    U_{\rm vol}(\u,p,J) :=& 
    \int_{V_0}p\,J\,\ii:\nabla\,\u\,dV_0 + U_{\rm vol}(\u_0,p_0,J_0),\\
    U_{\rm iso}(\u) :=& 
    \int_{V_0}\btau_{\rm iso}:\nabla\,\u\,dV_0 + U_{\rm iso}(\u_0,p_0,J_0),\\
    U_{\rm mus,base}(\u) :=&\, 
    2\int_{V_0}\sigma_{\rm mus,base}(I_1)\,\bb\,\frac{\partial I_1}{\partial\bb}:\nabla\,\u\,dV_0 + U_{\rm mus,base}(\u_0,p_0,J_0),\\
    U_{\rm apo,base}(\u) :=&\, 
    2\int_{V_0}\sigma_{\rm apo,base}(I_1)\,\bb\,\frac{\partial I_1}{\partial\bb}:\nabla\,\u\,dV_0 + U_{\rm apo,base}(\u_0,p_0,J_0),\\
    U_{\rm act}(\u) :=&\, 
    \sigma_0\int_{V_0}\frac{\hat{a}(q_0,t)}{\lambda_{\rm iso}}\sigma_{\rm act}(\lambda_{\rm iso})\,\bb\,\frac{\partial \lambda_{\rm iso}}{\partial\bb}:\nabla\,\u\,dV_0 + U_{\rm act}(\u_0,p_0,J_0),\\
    U_{\rm pas}(\u) :=&\, 
    \sigma_0\int_{V_0}\frac{\sigma_{\rm pas}(\lambda_{\rm iso})}{\lambda_{\rm iso}}\,\bb\,\frac{\partial \lambda_{\rm iso}}{\partial\bb}:\nabla\,\u\,dV_0, + U_{\rm pas}(\u_0,p_0,J_0)\\
    U_{\rm apo,fibre}(\u) :=& 
    \,\sigma_0\int_{V_0}\frac{\sigma_{\rm apo,fibre}(\lambda_{\rm iso})}{\lambda_{\rm iso}}\,\bb\,\frac{\partial \lambda_{\rm iso}}{\partial\bb}:\nabla\,\u\,dV_0 + U_{\rm apo,fibre}(\u_0,p_0,J_0).
\end{align*}

We assume that $\u_0$, $p_0$ and $J_0$ are the displacement, pressure and dilation in $V_0$. In our simulations we
set the displacement $\u_0$ to be the zero vector $\zero$, the pressure $p_0$ to be zero, and the dilation $J_0$ to be 1. The definition of the intrinsic properties of the tissues implies that all the strain-energy potentials in
the equations above at the state $(\u_0,p_0,J_0)$ vanish.

\section{Experimental measurements from MRI and DTI}
This appendix describes experimental data collection and analysis for measurement of the
three-dimensional changes in whole-muscle shape, and muscle fibre orientation of the human medial
gastrocnemius during fixed-end (constant ankle angle) plantarflexion contractions.

\subsection{Data acquisition}\label{appendix:data}
We obtained mDixon magnetic resonance imaging (MRI) and diffusion tensor imaging (DTI)
scans of the right lower legs of four female participants (age 29 $\pm$ 4 years mean $\pm$ standard deviation). All
procedures conformed to the Declaration of Helsinki (2008) and were approved by University of
New South Wales’ Human Research Ethics Committee HREC (approval HC17106). Informed
consent was obtained from all individual participants included in the study.
We instructed participants to lie supine with the knee slightly flexed in a 3T MRI scanner
(Philips Achieva TX, Best, The Netherlands). Their knee was supported by a foam wedge to maintain
a small gap between the MRI table and the posterior calf. Their right foot was strapped tightly into a
footplate, which was connected to a custom-built MRI-compatible force transducer. Their ankle was
held in 5 degrees plantarflexion relative to the neutral position with the foot perpendicular to the tibia. MRI
and DTI scans were obtained while the participant’s muscles were relaxed (rest) and during
plantarflexion contractions at 10\% (twice) and 20\% (once) of their maximum voluntary isometric
plantarflexion torque, which was previously determined with a dynamometer. We chose scan
sequences and fields of view to have a maximal scan duration of 2.5 minutes, because pilot testing
showed that all participants could maintain 20\% contractions for up to 2.5 minutes while staying
sufficiently still to obtain high quality MRI scans. mDixon and DTI scans were obtained during
separate contractions with participants given at least two minutes rest in between scans. The scans
did not cover the most proximal part ($\sim5$ cm) of the medial gastrocnemius. Visual feedback was
provided on a monitor next to the MRI bed to help participants maintain constant force during the
scans.

The settings of the mDixon scan were: 2-point 3D mDixon FFE, TR/TE1/TE2 6/3.5/4.6 msec,
field of view 180x180 mm, 250 slices, slice thickness 1 mm, acquisition matrix 180x180
(reconstructed to 192x192), reconstructed voxel size 0.94x0.94x1 mm, number of signal averages 1
and scan time 138 sec. The settings of the DTI scan were: DT-EPI with spectral pre-saturation with
inversion recovery (SPIR) fat suppression, TR/TE 8715/63 msec, field of view 180x180 mm, 40
slices, slice thickness 5 mm, 16 gradient directions on a hemisphere, number of signal averages
(NSA) 2, b = 500 s/mm$^2$ ($b0$ image with $b = 0$ s/mm$^2$ ), diffusion gradient time $\Delta/\delta = 30.4/8.2$ msec
and scan time 134 sec.

\subsection{Medial gastrocnemius 3D surface models}\label{appendix:gastroc}
We created 3D surface models of the medial gastrocnemius muscle at rest by manually
outlining the boundary of the medial gastrocnemius on the water image of the mDixon scan
(Bolsterlee et al., 2017) (Figure 2). We used non-rigid registration algorithms (Elastix 4.7) to create
surface models of the deformed muscles during contractions and visually inspected the results to
determine that the deformed outlines followed closely the boundaries of the muscle on the scans
obtained during contractions. The surface models were rotated to a local coordinate system using
principal component analysis on the vertices (nodes) of the surface model so that the x-axis aligns
with the long axis and the y- and z-axis with the width and depth axes, respectively.

\subsection{Fibre orientation at rest and during contractions}\label{appendix:fibre}
We determined muscle fibre orientations from the DTI scans (Bolsterlee et al., 2019) to
obtain maps with directions of the primary eigenvector of the diffusion tensor, i.e. maps of muscle
fibre orientations (Damon et al., 2002). The fibre orientations (pennation angles) were calculated as
the angle between the fibre orientations and the x-axis.

\paragraph{Conflict of Interest} The authors declare that the research was conducted in the absence of any commercial or financial
relationships that could be construed as a potential conflict of interest.

\paragraph{Author Contributions} JW, SR, DR, SD, NN contributed to the study design. JW, SR, DR, RK, SD, NN contributed to the
model development. BB conducted the MRI measurements and analysis. SR and BB ran all the
simulations for the paper. JW, SR, DR, BB, RK, SD, NN contributed to the data analysis and
manuscript preparation.

\paragraph{Funding} We thank the Natural Sciences and Engineering Research Council of Canada for Discovery Grants to JW and NN, an Alexander Graham Bell Canada Graduate Scholarship-Doctoral to SR, and an
Undergraduate Student Research Award to RK. We are also grateful for funding to SD from
Comisión Nacional de Investigación Científica y Tecnológica of Chile through Becas-Chile.

\section*{Figures}
\begin{figure}[!ht]
    \centering
    \includegraphics[width = 1\textwidth, 
height=.75\textheight]{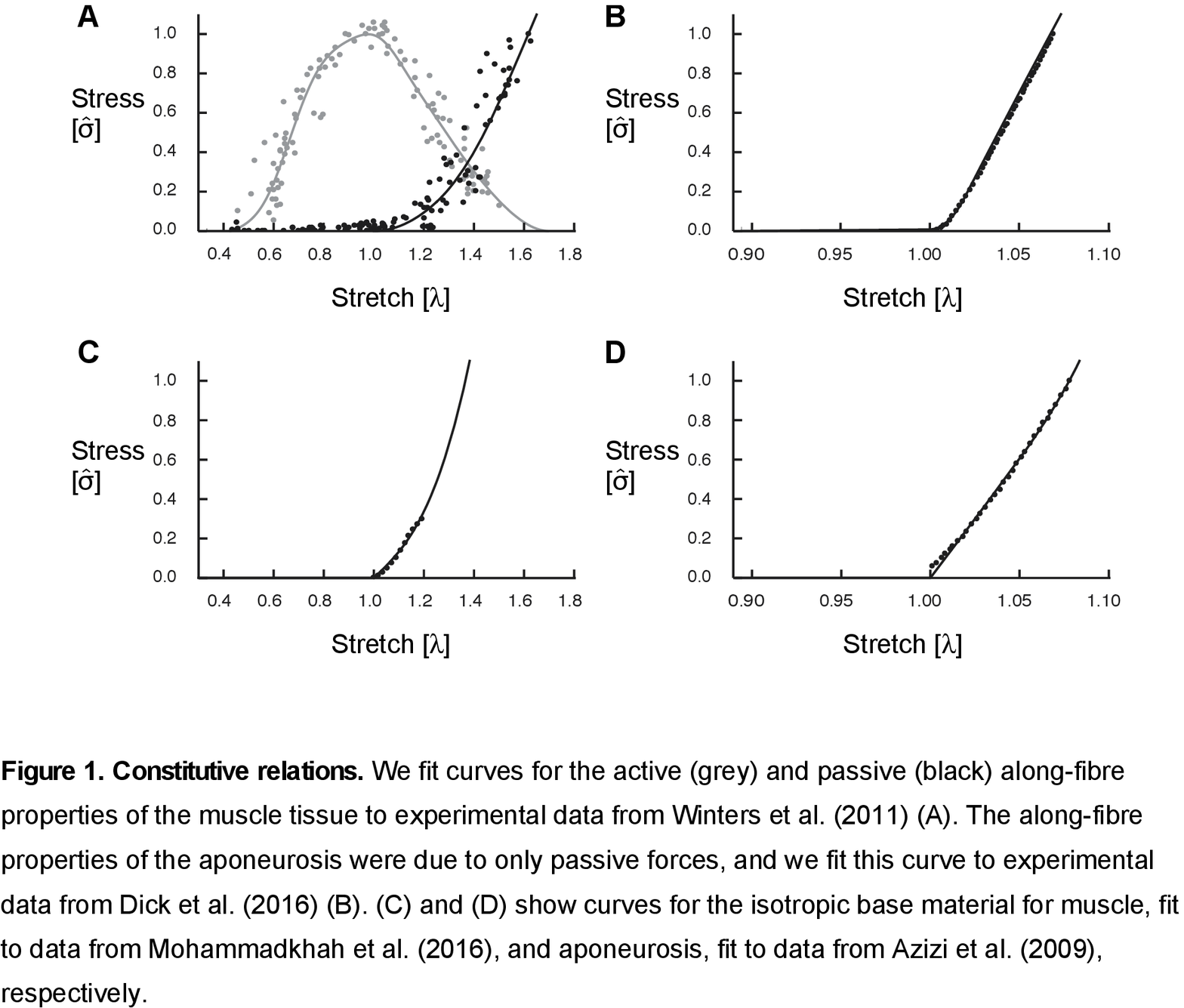}
\end{figure}
\begin{figure}[!ht]
    \centering
    \includegraphics[width = 1\textwidth, 
height=.8\textheight]{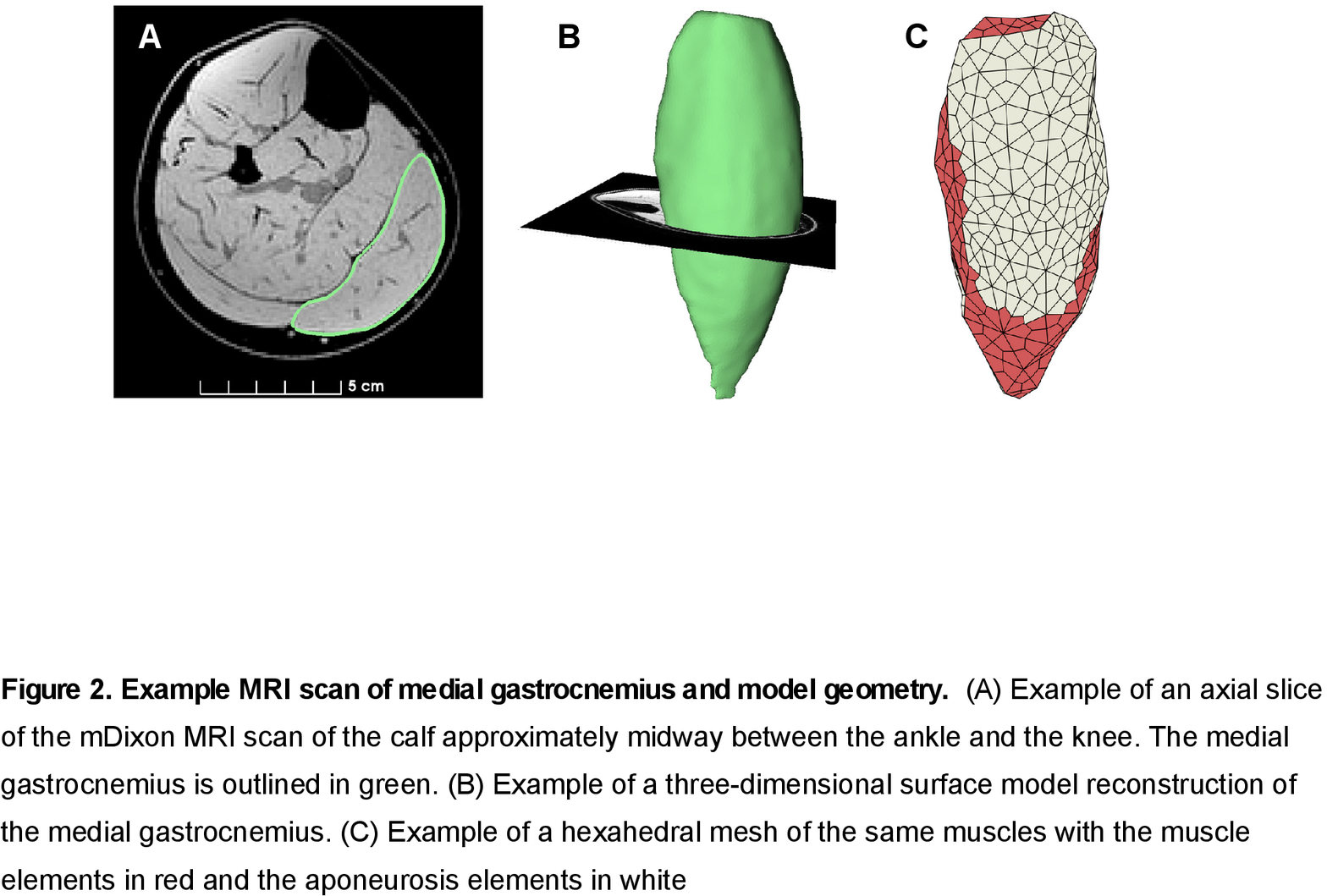}
\end{figure}
\begin{figure}[!ht]
    \centering
    \includegraphics[width = 1\textwidth, 
height=1\textheight]{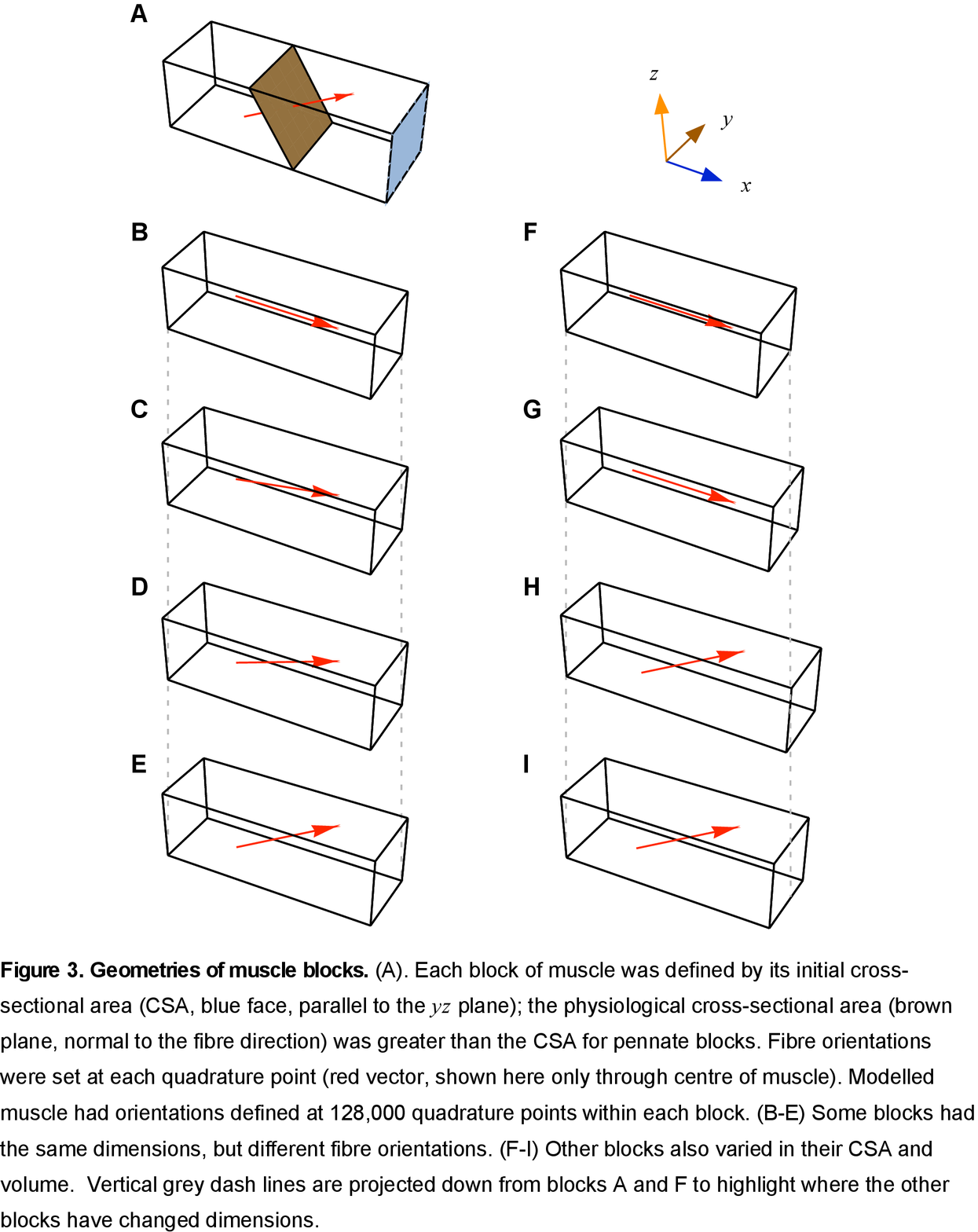}
\end{figure}
\begin{figure}[!ht]
    \centering
    \includegraphics[width = 1\textwidth, 
height=1\textheight]{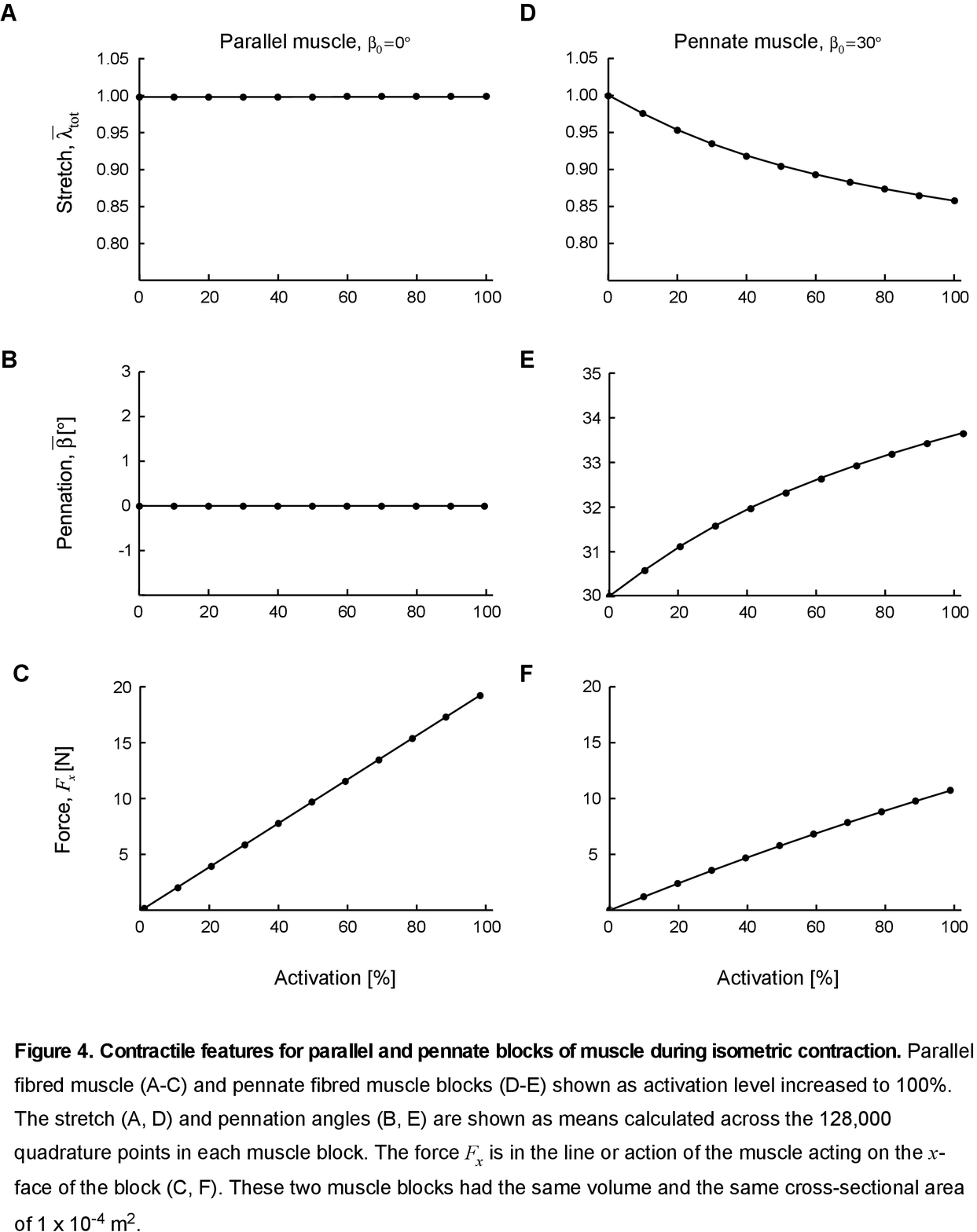}
\end{figure}
\begin{figure}[!ht]
    \centering
    \includegraphics[width = 1\textwidth, 
height=1\textheight]{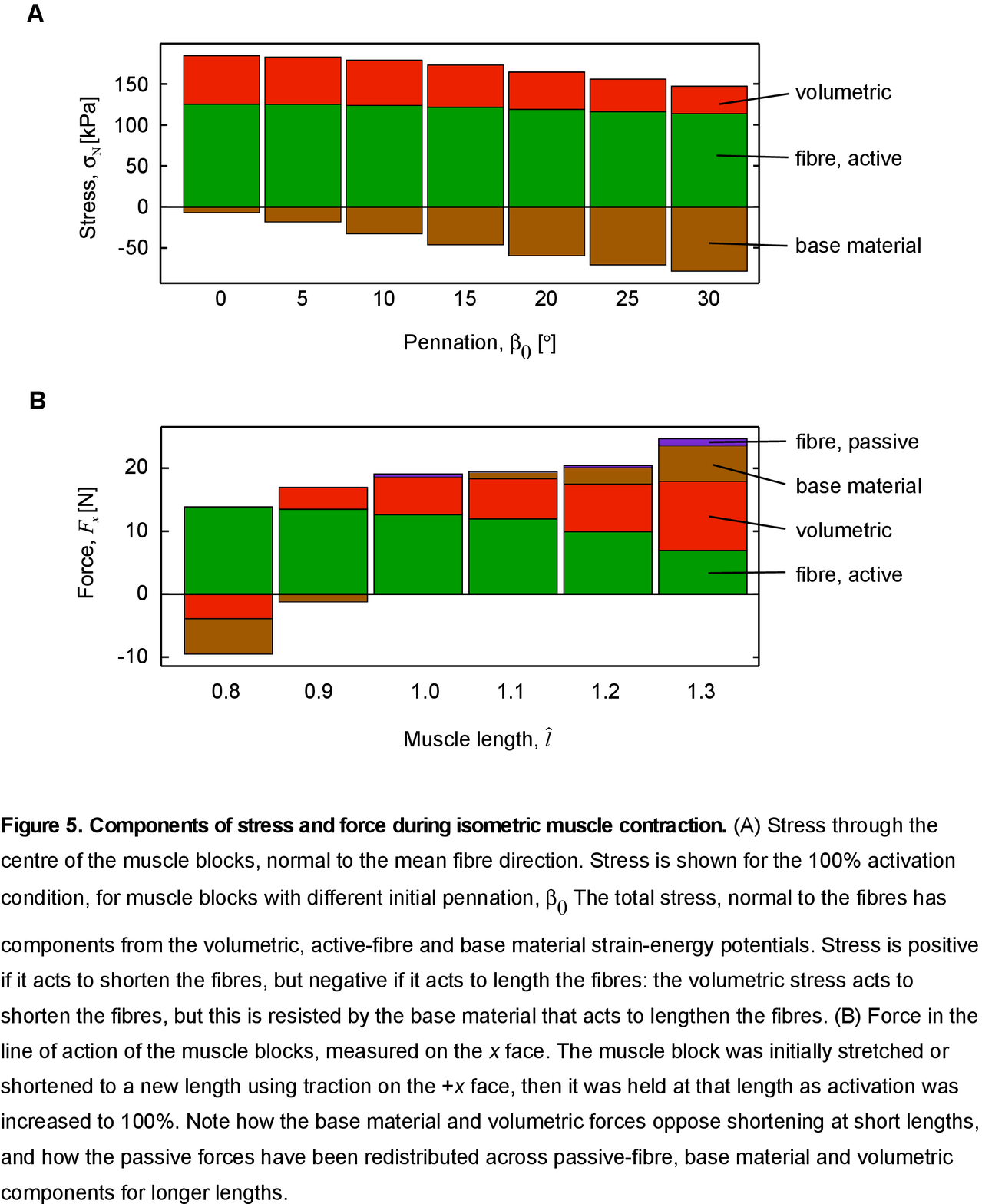}
\end{figure}
\begin{figure}[!ht]
    \centering
    \includegraphics[width = 1\textwidth, 
height=1\textheight]{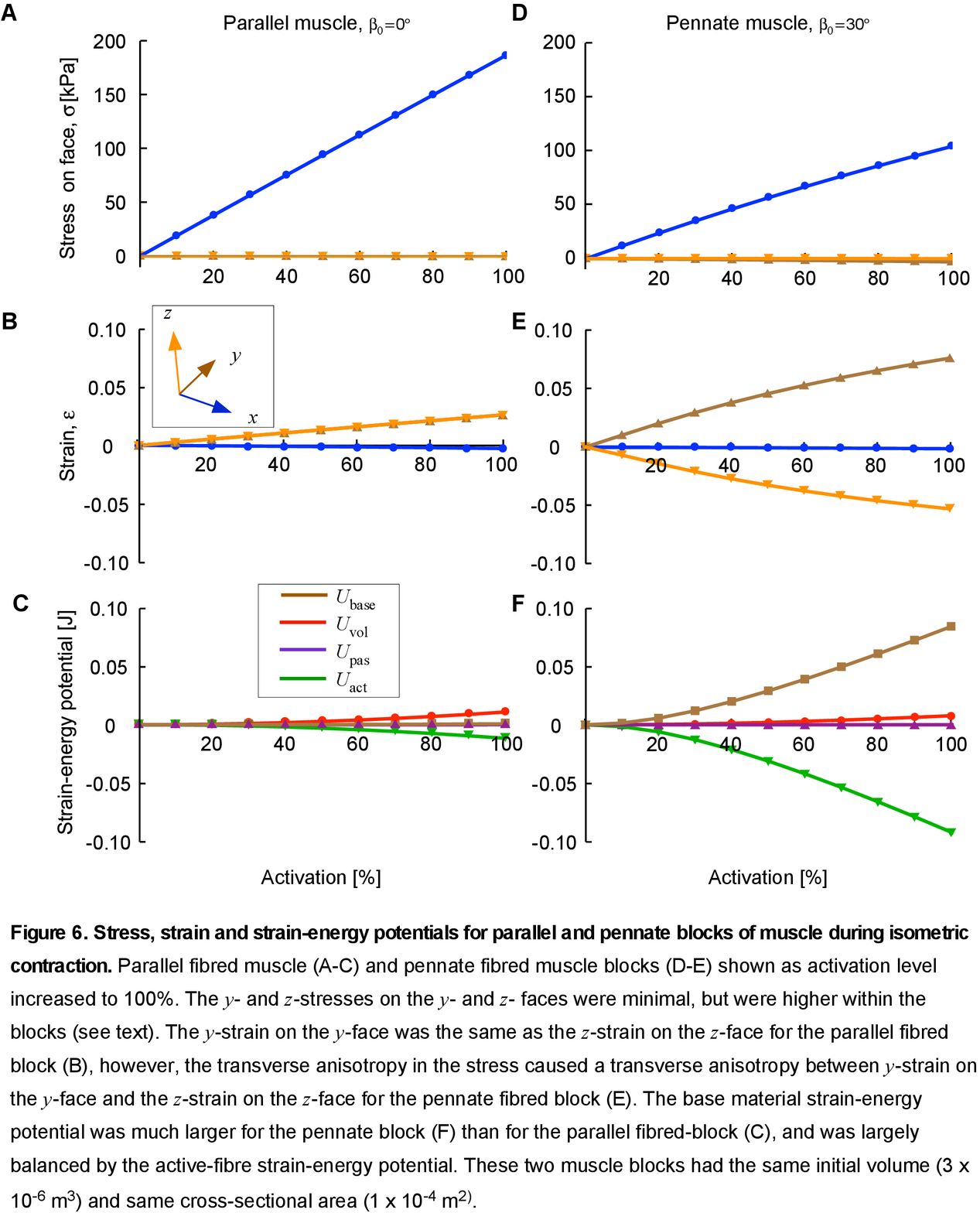}
\end{figure}
\begin{figure}[!ht]
    \centering
    \includegraphics[width = 1\textwidth, 
height=.8\textheight]{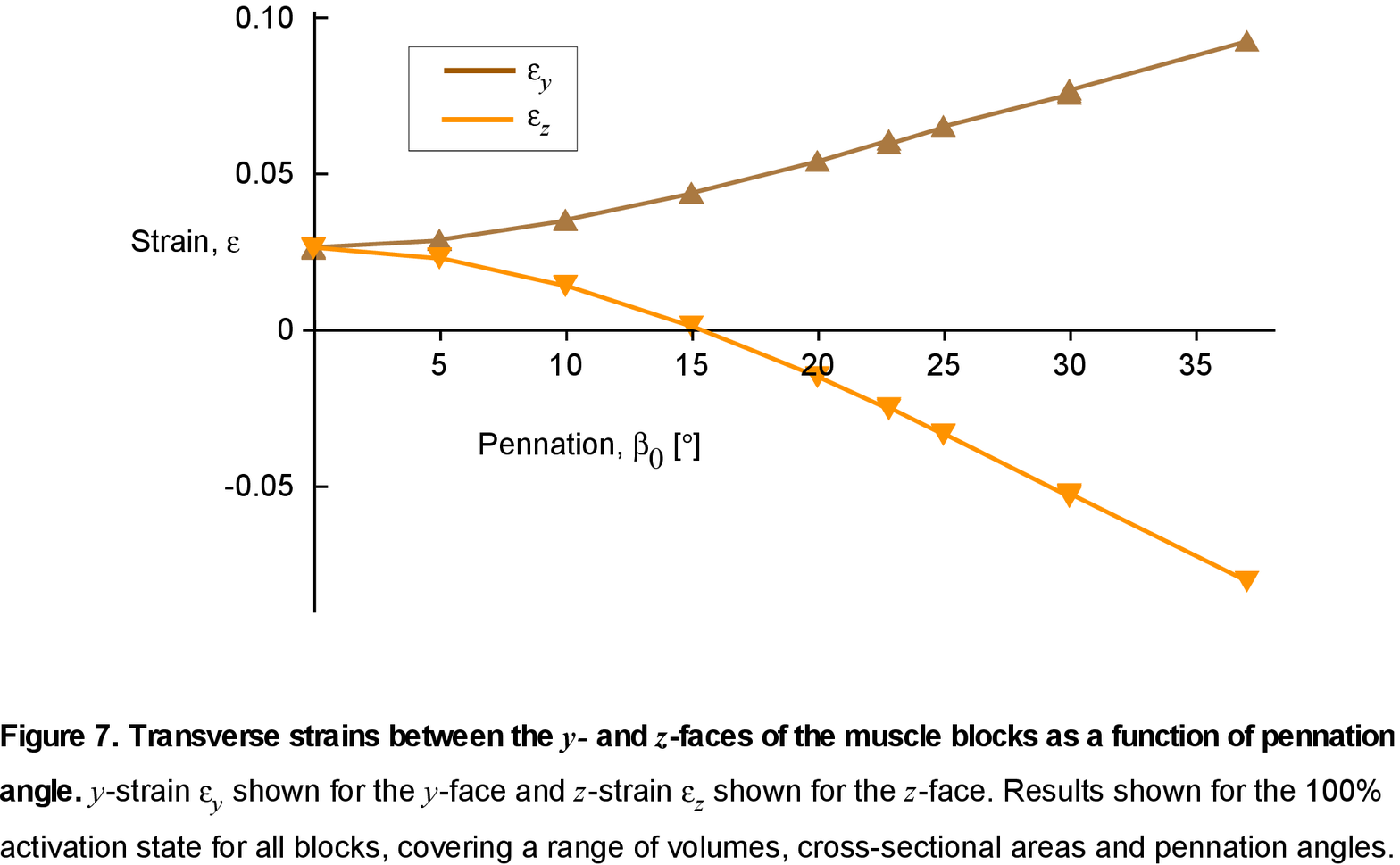}
\end{figure}
\begin{figure}[!ht]
    \centering
    \includegraphics[width = 1\textwidth, 
height=1\textheight]{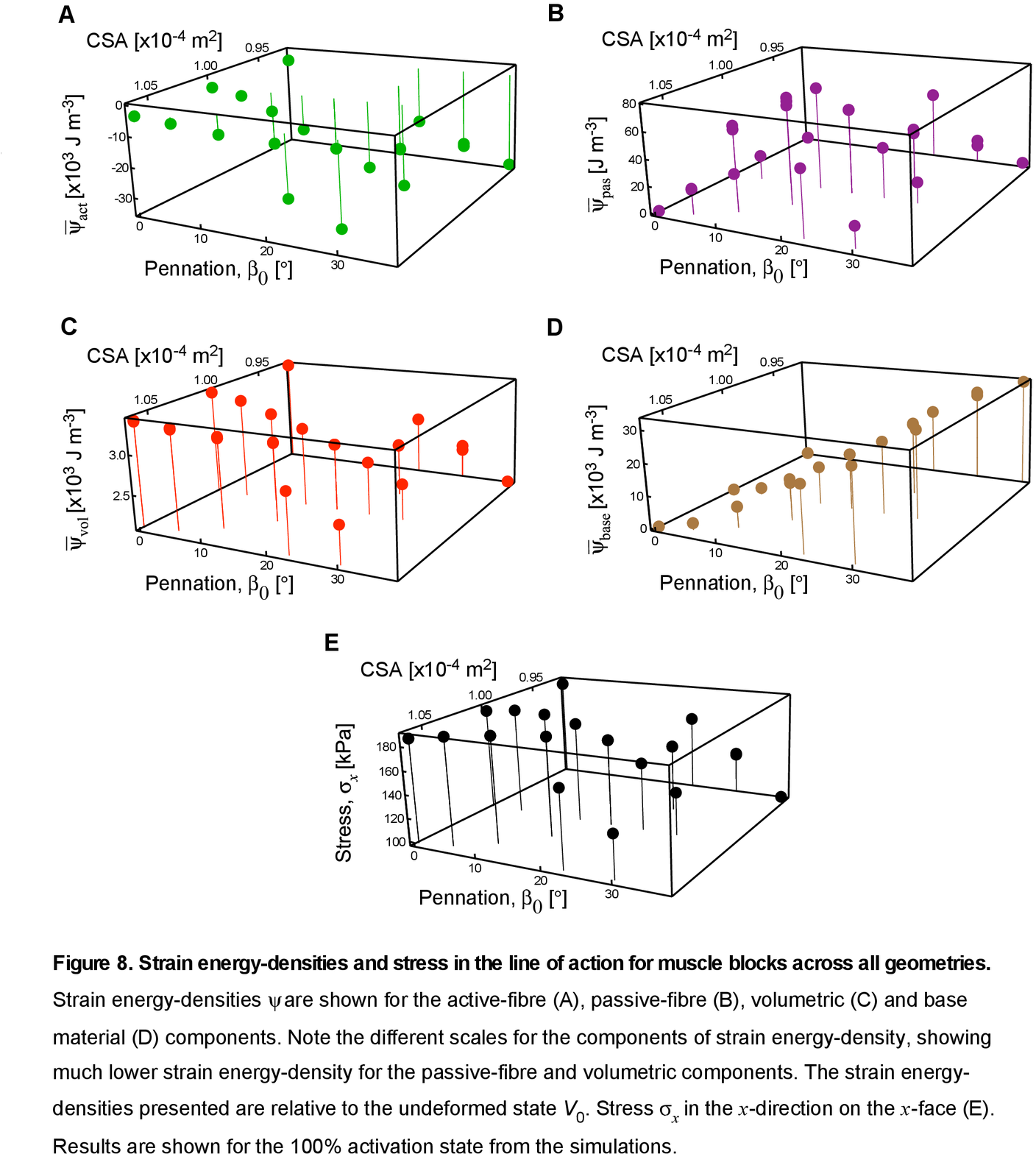}
\end{figure}
\begin{figure}[!ht]
    \centering
    \includegraphics[width = 1\textwidth, 
height=1\textheight]{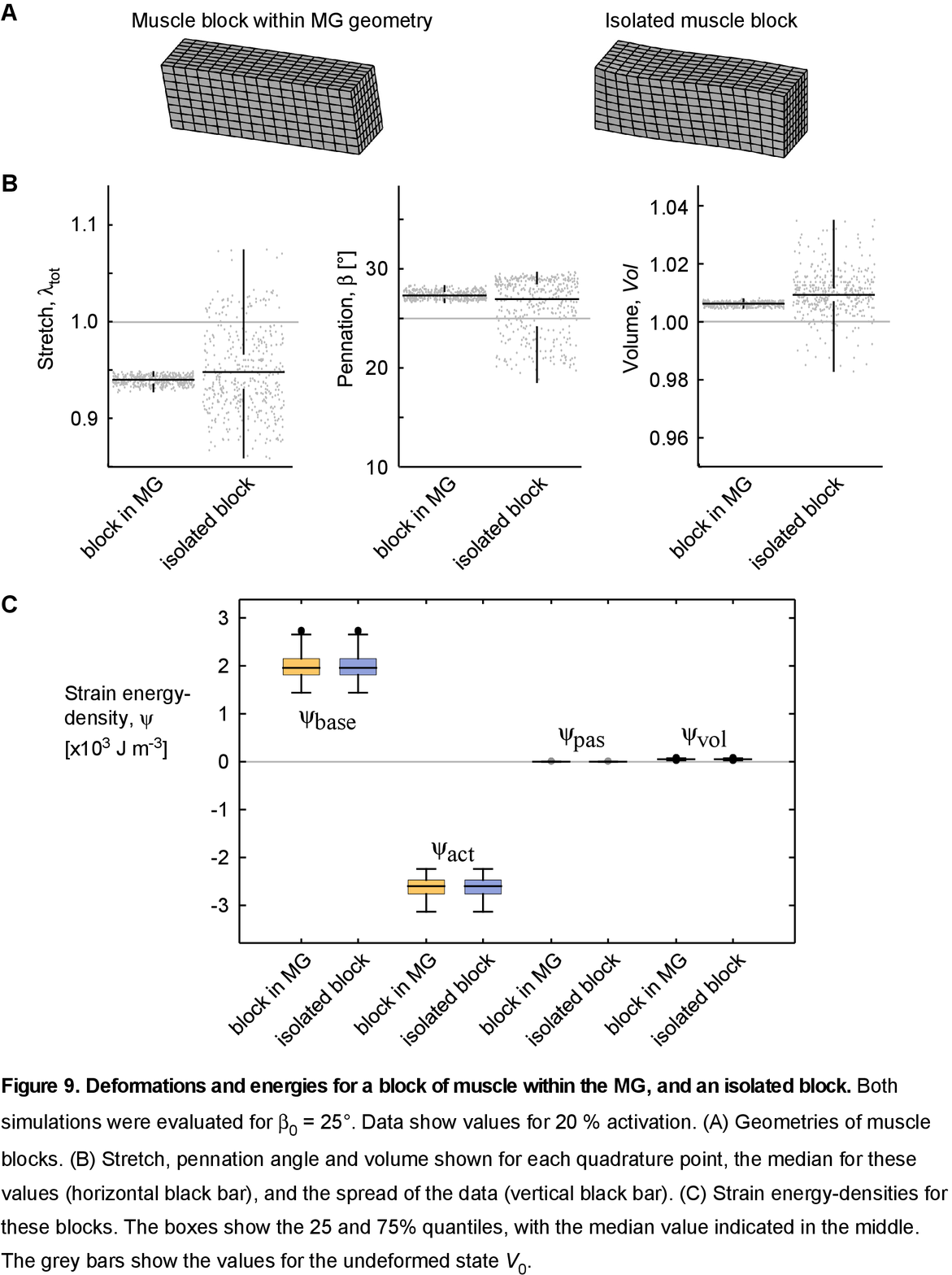}
\end{figure}
\begin{figure}[!ht]
    \centering
    \includegraphics[width = 1\textwidth, 
height=.9\textheight]{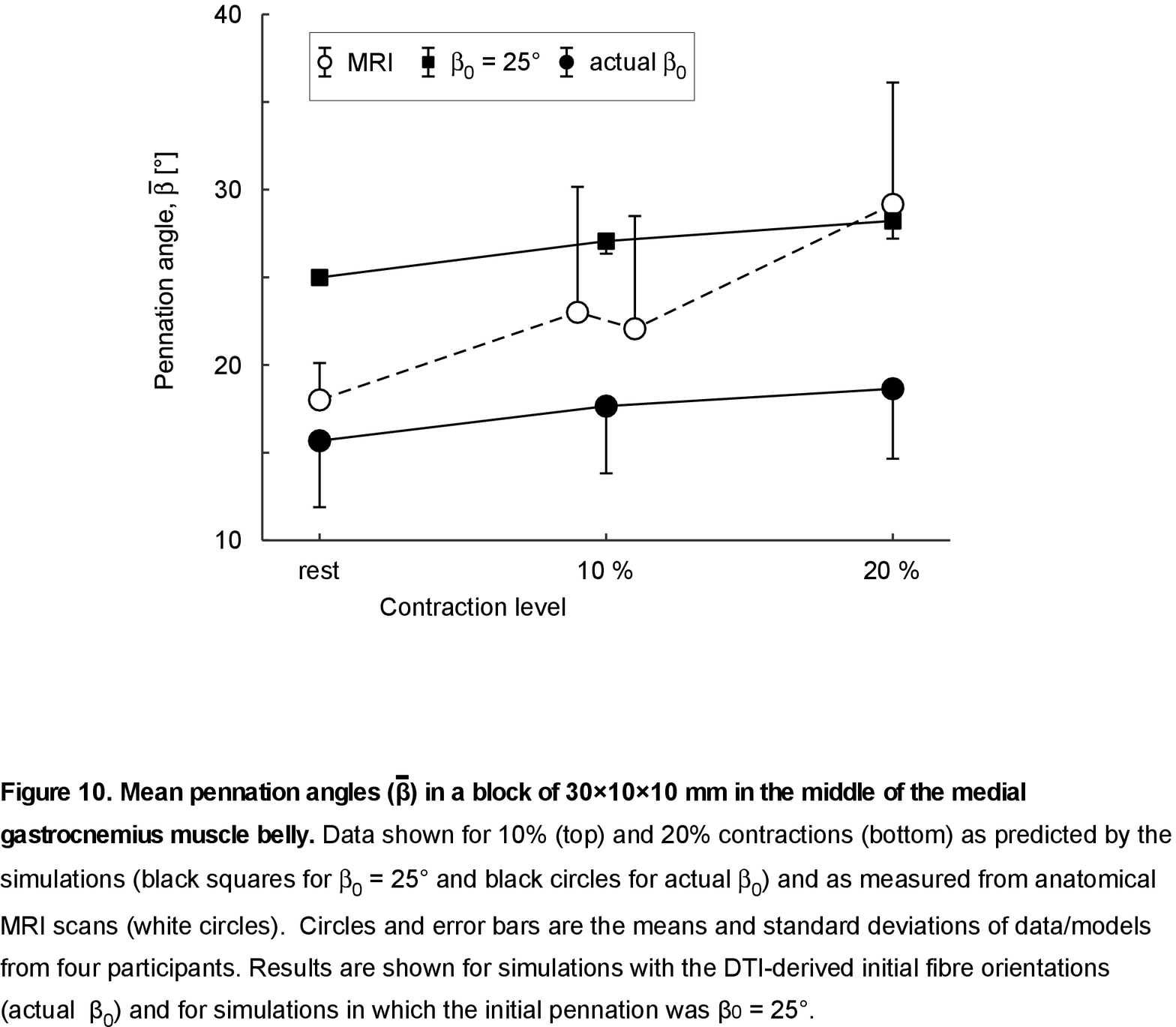}
\end{figure}
\begin{figure}[!ht]
    \centering
    \includegraphics[width = 1\textwidth, 
height=1\textheight]{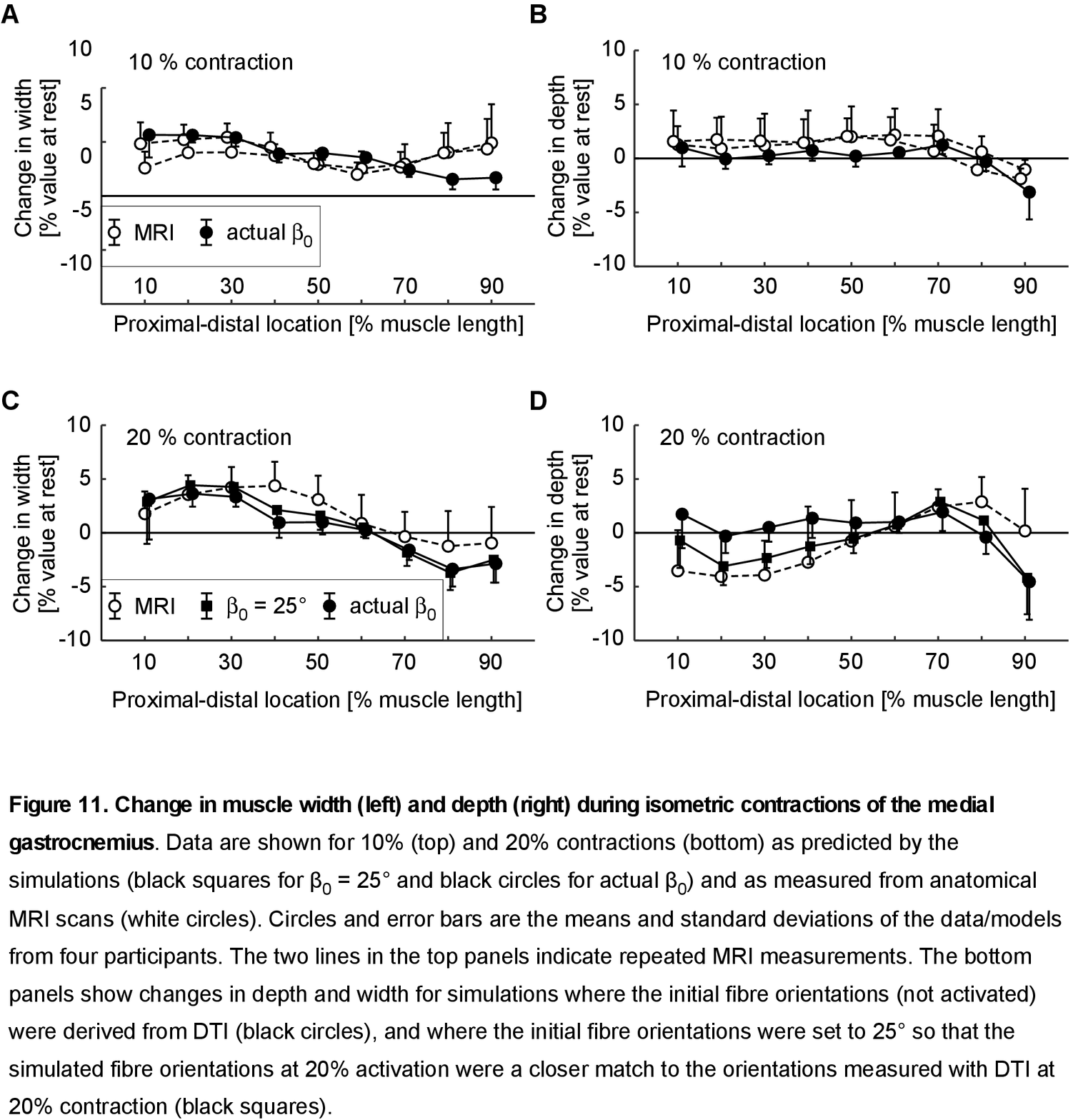}
\end{figure}
\begin{figure}[!ht]
    \centering
    \includegraphics[width = 1\textwidth, 
height=1\textheight]{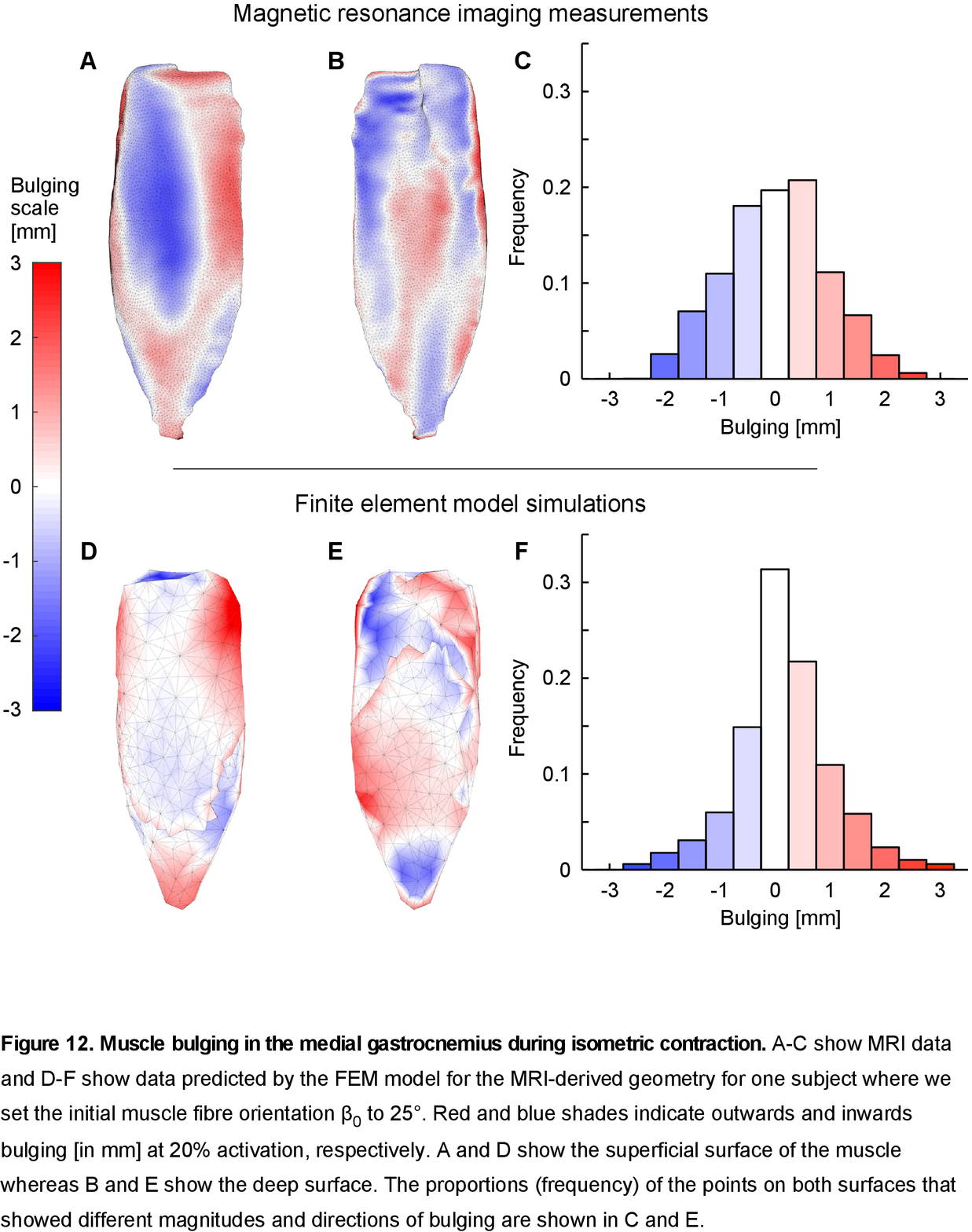}
\end{figure}

\clearpage
\section*{Tables}
\begin{table}[ht!]
{\centering\begin{tabular}{|l|l|}
\hline
{\bf Symbol}                    & {\bf Definition}                                        \\ \hline
$\u$                      & displacement vector                               \\ \hline
$p$                       & internal pressure                                 \\ \hline
$J$                       & dilation                                          \\ \hline
$E_{\rm tot}$             & total strain energy                               \\ \hline
$U$                       & strain-energy potential                           \\ \hline
$U_{\rm int}$             & internal strain-energy potential                  \\ \hline
$W_{\rm ext}$             & work done by external forces                      \\ \hline
$DE_{\rm tot}$            & first variation of $E_{\rm tot}$                  \\ \hline
$\Psi$                    & strain energy-density                             \\ \hline
$V_0$                     & initial configuration                             \\ \hline
$V$                       & current configuration                             \\ \hline
CSA                       & cross-sectional area                              \\ \hline
$\beta_0$                 & pennation angle in initial configuration $V_0$    \\ \hline
$\beta$                   & pennation angle in current configuration $V$      \\ \hline
$\hat\beta$               & Mean pennation angle in current configuration $V$ \\ \hline
$Vol$                     & current volume                                    \\ \hline
$\lambda_{\rm iso}$       & isovolumetric stretch                             \\ \hline
$\bar{\lambda}_{\rm iso}$ & mean total stretch                                \\ \hline
$s_base$                  & stiffness parameter for muscle base material      \\ \hline
$\varepsilon$             & scalar strain                                     \\ \hline
$l$                       & length                                            \\ \hline
$\hat{l}$                 & normalized length                                 \\ \hline
$F$                       & force                                             \\ \hline
$F_x$                     & force in x-direction                              \\ \hline
$\sigma$                  & scalar stress                                     \\ \hline
$\kappa$                  & bulk modulus                                      \\ \hline
\end{tabular}\caption{Symbols and definitions of variables in the main text}}
\end{table}

\begin{table}[ht!]
{\centering\resizebox{\textwidth}{!}{\begin{tabular}{|l|l|l|l|}
\hline
{\bf Symbol} & {\bf Definition} & {\bf Symbol} & {\bf Definition} \\\hline
$S_0$ 	&	 surface of $V_0$ 	&	$\hat{\n}_0$ 	&	normal unit vector on $S_0$	  \\\hline
$V $ 	&	 current configuration  &	$\hat{\n}$ 	&	normal unit vector on $S$ \\\hline
$S$ 	&	 surface of $V$	  	&	$\hat{\a}_0$ 	&	 normalized fibre orientation in $V_0$ \\\hline
$S_{\rm t}$ 	&	 region of $S$ with applied traction	&	 $\t$ 	&	 applied traction \\\hline
$S_{\rm d}$ 	&	 region of $S$ with applied displacement 	& 	$\kappa_{\rm mus}$ 	&	 bulk modulus muscle	  \\\hline
${\rm Vol}$ 	&	 current volume 	&	$\kappa_{\rm apo}$ 	&	 bulk modulus aponeurosis\\\hline
$A$ 	&	 area in $V$ 	&	$\bb$ 	&	 left Cauchy tensor \\\hline
$t$ 	&	 time 	&	$\bb_{\rm iso}$ 	&	 isovolumetric left Cauchy tensor \\\hline
$q_0$ 	&	 point in $V_0$ 	&	$\bsigma$ 	&	 Cauchy stress tensor \\\hline
$q $ 	&	 point in $V $ 	&	$\btau$ 	&	 Kirchhoff tensor \\\hline
$U_{\rm vol}$ 	&	 volume energy potential 	&	$I_1$ 	&	 first invariant \\\hline
$U_{\rm iso}$ 	&	 isovolumetric energy potential 	&	 $I_3$ 	&	 third invariant\\\hline
$U_{\rm base}$ 	&	 base material energy potential 	&	 $I_4$ 	&	 fourth invariant \\\hline
$U_{\rm fibre}$ 	&	 fibre energy potential 	&	 	$\hat{a}$ 	&	activation level in muscle fibres	  \\\hline
$U_{\rm apo,base}$ 	&	 aponeurosis base material energy potential	&	 $\sigma_0$ 	&	 maximum isometric stress of contractile elements \\ \hline
$U_{\rm apo,fibre}$ 	&	 aponeurosis fibre energy potential 	&	 $\ff$	&	deformation tensor  \\\hline
$U_{\rm mus,base}$ 	&	 muscle base material energy potential 	&   $\ii$	&	identity tensor \\\hline
$U_{\rm mus,fibre}$ 	&	 muscle fibre energy potential 	&	$\sigma_{\rm base}$ 	&	 base material stress \\\hline
$U_{\rm act}$ 	&	 active muscle fibre energy potential 	&	$\sigma_{\rm mus,base}$ 	&	 muscle base material stress \\\hline
$U_{\rm pas}$ 	&	passive muscle fibre energy potential 	&	$\sigma_{\rm apo,base}$ 	&	 aponeurosis base material stress \\\hline
$\Psi_{\rm vol}$ 	&	 volume energy-density	&	 $\sigma_{\rm fibre}$ 	&	 fibre stress \\\hline
$\Psi_{\rm iso}$ 	&	 isovolumetric energy-density	&	 $\sigma_{\rm mus,fibre}$ 	&	 muscle fibre stress \\\hline
$\Psi_{\rm base}$ 	&	 base energy-density	&	 $\sigma_{\rm apo,fibre}$ 	&	 aponeurosis fibre stress \\\hline
$\Psi_{\rm fibre}$ 	&	 fibre energy-density	&	 $\hat{\sigma}_{\rm act}$ 	&	 active muscle fibre stress \\\hline
$\Psi_{\rm apo,base}$ 	&	aponeurosis base material energy-density	&	 $\hat{\sigma}_{\rm pas}$ 	&	 passive muscle fibre stress \\\hline
$\Psi_{\rm apo,fibre}$ 	&	 aponeurosis fibre energy-density	&  $\nabla_0$	 & gradient with respect to $V_0$ \\\hline
$\Psi_{\rm mus,base}$ 	&	muscle base material energy-density	&	$\nabla$ 	&	gradient with respect to $V$  \\\hline
$\Psi_{\rm mus,fibre}$ 	&	muscle fibre energy-density	&	$\bdiv$ 	&  tensorial divergence with respect to $V$ \\\hline
&   &  $\zero$ & zero vector \\\hline
\end{tabular}}\caption{Variables only in the appendix}}
\end{table}

\clearpage
\nocite{*}
\bibliographystyle{apalike}
\bibliography{paper1.bib}

\end{document}